\documentclass[pra,onecolumn,preprint,11pt,showpacs,citeautoscript,footinbib,eqsecnum]{revtex4-2}
\usepackage[T1]{fontenc}
\usepackage[utf8]{inputenc}
\usepackage{lipsum}
\usepackage{amsmath,bm}
\usepackage{amssymb}
\usepackage{yfonts}
\usepackage{eufrak}
\usepackage{bbm}
\usepackage{braket}
\usepackage{graphicx}
\usepackage{xcolor}
\usepackage{pifont}
\usepackage[mathscr]{euscript}
\usepackage[shortlabels]{enumitem}
\usepackage[justification=justified]{subcaption}
\usepackage{float}
\usepackage{dsfont}
\usepackage{comment}
\usepackage{subcaption}
\usepackage{slashed}
\usepackage[normalem]{ulem} 

\usepackage[papersize={8.5in,11in}]{geometry}
\usepackage[colorlinks=true]{hyperref}  
\hypersetup{
    unicode=false,          
    pdftoolbar=true,        
    pdfmenubar=true,        
    pdffitwindow=false,     
    pdfstartview={FitH},    
    pdftitle={Flstar},    
    pdfauthor={Christos, Sachdev, Luo},     
    pdfsubject={},   
    pdfcreator={},   
    pdfproducer={}, 
    pdfkeywords={} {} {}, 
    pdfnewwindow=true,      
    colorlinks=true,       
    linkcolor=blue, 
    citecolor=blue,        
    filecolor=magenta,      
    urlcolor=blue           
} 
\renewcommand{\vec}[1]{\boldsymbol{#1}}

\newcommand{\up}{\uparrow}
\newcommand{\down}{\downarrow}

\newcommand{\bQ}{\mathbf{Q}}
\newcommand{\bk}{\mathbf{k}}
\newcommand{\bR}{\mathbf{R}}
\newcommand{\dmu}{\delta\mu}
\newcommand{\avH}[1]{\langle \vec{H}_{#1}\rangle}

\usepackage{xcolor}
\definecolor{brandeisblue}{rgb}{0.0, 0.44, 1.0}

\usepackage{mathtools} 
\begin{document}
\graphicspath{{figures/}}

\title{Canted magnetism and $\mathbb{Z}_2$ fractionalization in metallic states\\ of the Lieb lattice Hubbard model near quarter filling}

\begin{abstract} 
A recent experiment has examined ultracold, fermionic, spin-1/2 $^6$Li atoms in the Lieb lattice at different Hubbard repulsion $U$ and filling fractions $\nu$ (Lebrat {\it et al.\/} \href{https://arxiv.org/abs/2404.17555}{arXiv:2404.17555}). At $\nu=1/2$ and small $U$, they observe an enhanced compressibility on the $p_{x,y}$ sites, pointing to  
a flat band near the Fermi energy. At $\nu=1/2$ and large $U$ they observe an insulating ferrimagnet. Both small and large $U$ observations at $\nu=1/2$ are consistent with theoretical expectations. 
Surprisingly, near $\nu=1/4$ and large $U$, they again observe a large $p_{x,y}$ compressibility, pointing to a flat $p_{x,y}$ band of fermions across the Fermi energy. Our Hartree-Fock computations near $\nu=1/4$ find states with canted magnetism (and related spiral states) at large $U$, which possess nearly flat $p_{x,y}$ bands near the Fermi level. We employ parton theories to describe quantum fluctuations of the magnetic order found in Hartree-Fock. We find a metallic state with $\mathbb{Z}_2$ fractionalization possessing gapless, fermionic, spinless `chargons' carrying $\mathbb{Z}_2$ gauge charges which have a nearly flat $p_{x,y}$ band near their Fermi level: this fractionalized metal is also consistent with observations. Our DMRG study does not indicate the presence of magnetic order, and so supports a fractionalized ground state. Given the conventional ferrimagnetic insulator at $\nu=1/2$, the $\mathbb{Z}_2$ fractionalized metal at $\nu=1/4$ represents a remarkable realization of doping-induced fractionalization.
\end{abstract}

\author{Alexander Nikolaenko}
\thanks{These authors contributed equally to this work.}
\affiliation{Department of Physics, Harvard University, Cambridge MA 02138, USA}

\author{Pietro M. Bonetti}
\thanks{These authors contributed equally to this work.}
\affiliation{Department of Physics, Harvard University, Cambridge MA 02138, USA}

\author{Anant Kale}
\affiliation{Department of Physics, Harvard University, Cambridge MA 02138, USA}

\author{Martin Lebrat}
\affiliation{Department of Physics, Harvard University, Cambridge MA 02138, USA}

\author{Markus Greiner}
\affiliation{Department of Physics, Harvard University, Cambridge MA 02138, USA}

\author{Subir Sachdev}
\affiliation{Department of Physics, Harvard University, Cambridge MA 02138, USA}

\maketitle
\newpage
\linespread{1.05}
\tableofcontents

\section{Introduction}
\label{sec:intro}

The {\it square lattice\/} Hubbard model of spin-1/2 fermions plays a fundamental role in the theory of cuprate high temperature superconductivity. Although at filling fraction $\nu=1/2$ and large Hubbard repulsion $U$ this model has long-range antiferromagnetic order, fractionalized spin liquids have been proposed \cite{Anderson87,LeeWenRMP} to play a role in the physics upon hole doping away from $\nu=1/2$. In recent studies \cite{Christos23,SSZaanen}, fractionalization is important in the intermediate temperature pseudogap regime, but the lowest temperature state is a conventional state in which the fractionalized excitations confine. The present paper will examine the theory of the Hubbard model on the {\it Lieb lattice\/}. Here too, the large $U$ insulator at $\nu=1/2$ is conventional, with long-range ferrimagnetic order. But we propose that the Lieb lattice can realize the remarkable phenomenon of doping-induced fractionalization in the metallic low temperature state near $\nu = 1/4$.

\begin{figure}[t!]
    \centering
    \includegraphics[width=0.5\linewidth]{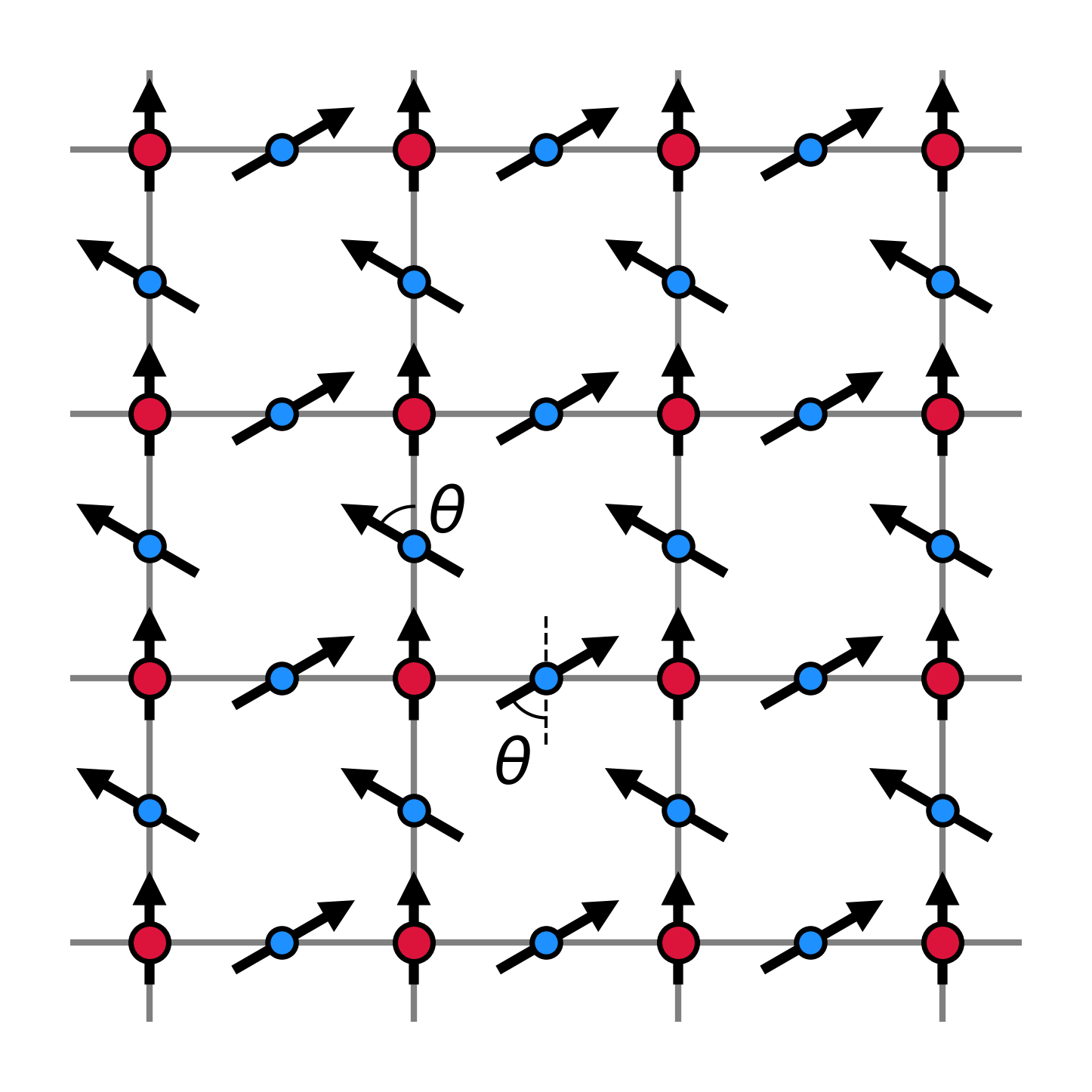}
    \caption{Spin pattern of the \textit{canted} magnetic state obtained at quarter filling within Hartree-Fock theory. The magnetizations on $p_x$ and $p_y$ sites (marked in blue) are equal in absolute value but they form an opposite angle ($\theta$) with respect to the magnetization on $d$ sites (marked in red).}
    \label{fig:canted}
\end{figure}
Our theoretical work follows the observations of 
Lebrat {\it et al.} \cite{Lebrat24}, who studied ultracold, fermionic, spin-1/2 $^{6}$Li atoms in an optical lattice, realizing the Hubbard model on a Lieb lattice with nearest-neighbor hopping. Their ability to vary the filling fraction $\nu$ and the ratio of the Hubbard repulsion $U$ to the hopping $t$ has uncovered a rich phase diagram.
For $U=0$, the Lieb lattice features an exactly flat band in the middle of the fermionic spectrum (see Fig.~\ref{fig: Lieb lattice}), and a finite temperature signature of the flatband was observed as a large compressibility on the $p_{x,y}$ sites at half-filling. 
For large $U$, an insulating ferrimagnetic state is expected~\cite{Gouveia2015,Gouveia2016,Soni2021}
for $\nu=1/2$, and finite temperature signatures of the ferrimagnetic state were observed in the staggered magnetization and magnetic susceptibility.

The focus of the present paper is on the remarkable observations of Lebrat {\it et al.} \cite{Lebrat24} at large $U$ near $\nu = 1/4$. Here,
they again observed an enhanced compressibility on the $p_{x,y}$ sites, pointing to a metallic state of fermions with a nearly flat band near the Fermi level. This is surprising because the small $U$ free electron theory does not have any flat band near the Fermi level at $\nu=1/4$ (see Fig.~\ref{fig: Lieb lattice}).
A simple way to realize a $p_{x,y}$ flat band at $\nu=1/4$ at large $U$ is to assume that all spins are parallel, realizing a fully polarized ferromagnet of effectively spinless fermions at filling fraction $\nu_{\rm spinless} = 2 \nu = 1/2$. But the existing observations, and our numerical results presented here, do not support such a maximal polarization. Instead, our analysis here leads to two novel classes of metallic states for large $U$ near $\nu = 1/4$:
\begin{itemize}
\item Hartree-Fock computations reveal states with {\it canted\/} magnetism, and related spiral order, shown in Figs.~\ref{fig:canted} and \ref{fig:Fig2}.
The canted state is a partially polarized ferromagnet along one direction in spin space, and an antiferromagnet along an orthogonal direction. The fermionic dispersion of the canted state features a nearly-flat $p_{x,y}$ band near the Fermi level (see Figs.~\ref{fig:Fig3}, \ref{fig: bandstructures}). The Fermi level bands in the spiral states are not particularly flat (see Fig.~\ref{fig: spectrum spiral}).
\item We extend theories of quantum fluctuations of canted \cite{Wang16,CSS17} magnetism to obtain a gapless, metallic state with no net magnetization and spin rotation invariance and lattice symmetries preserved. This state features $\mathbb{Z}_2$ fractionalization \cite{NRSS91,Wen91,SS_QPM} with the same super-selection sectors (`anyons') as the toric code \cite{Kitaev03}.
In the $\epsilon$ sector we have gapless, spinless, fermionic `chargons' which contain a nearly-flat band near their Fermi level. In the $e$ sector, we have gapped, spinful, bosonic, spinons. Finally, there are also gapped, bosonic `visons' in the $m$ sector, which we will briefly note. Similar theories \cite{NRSS91,SSNR91} apply to the spiral states in Fig.~\ref{fig:Fig2}, and they also lead to $\mathbb{Z}_2$ fractionalization but with Ising-nematic order associated with breaking lattice 90$^\circ$ rotational symmetry. Insulating phases with $\mathbb{Z}_2$ or U(1) fractionalization on Lieb-like lattices with phonons have been considered in Refs.~\cite{Han2023,Han2024}.
\end{itemize}
The canted state is relatively easy to rule out by observing the absence of a net ferromagnetic moment.  The states with spiral magnetic order also have Ising-nematic order, and so would break lattice rotational symmetry in spin-insensitive observables. The numerical and theoretical results we present here favor the more exotic $\mathbb{Z}_2$ fractionalized state near $\nu = 1/4$. 

Our analysis begins in Section~\ref{sec:hf} by a Hartree-Fock analysis of the Hubbard model on Lieb lattice. This analysis leads to the phase diagram spiral, canted, and ferromagnetic states shown in Fig.~\ref{fig:Fig2} near $\nu = 1/4$. We also compute the band structure and site-resolved compressibilities of these states.

In Section~\ref{sec:su2} we describe quantum fluctuations of the canted and spiral states of the Hubbard model of Section~\ref{sec:hf} in terms of a SU(2) gauge theory \cite{sdw09,DCSS15b,CSS17,Scheurer:2017jcp,Sachdev:2018ddg,MSSS18,SSST19,ScheurerShen,Bonetti22}. This proceeds by expressing the fermions $c_\sigma$ ($\sigma = \uparrow, \downarrow$) of the Hubbard model in a rotating reference frame in spin space \cite{SS80,Schulz90,Dupuis02,Dupuis04}
\begin{align}
\left( \begin{array}{c}
c_\uparrow \\ c_\downarrow \end{array} \right) = \left( \begin{array}{cc} z_\uparrow & -z_\downarrow^\ast \\
z_\downarrow & z_\uparrow^\ast
\end{array} \right) \left( \begin{array}{c}
\psi_+ \\ \psi_- \end{array} \right)\,. \label{czpsi}
\end{align}
The $z_\sigma$ realize the bosonic spinons, and the $\psi_\pm$ are the fermionic chargons. The representation (\ref{czpsi}) introduces a SU(2) gauge symmetry under which 
\begin{align}
\left( \begin{array}{c}
\psi_+ \\ \psi_- \end{array} \right) \rightarrow U \left( \begin{array}{c}
\psi_+ \\ \psi_- \end{array} \right) \quad , \quad  \left( \begin{array}{cc} z_\uparrow & -z_\downarrow^\ast \\
z_\downarrow & z_\uparrow^\ast
\end{array} \right) \rightarrow  \left( \begin{array}{cc} z_\uparrow & -z_\downarrow^\ast \\
z_\downarrow & z_\uparrow^\ast
\end{array} \right) U^\dagger
\end{align}\label{eq: z gauge transformation}
where $U$ is an arbitrary SU(2) rotation, $U^\dagger U = 1$. It is important to note that this gauge rotation is distinct from (and commutes with) spin rotations $V$, which act as left multiplication on the $z$ matrix \cite{sdw09}
\begin{align}
    \left( \begin{array}{cc} z_\uparrow & -z_\downarrow^\ast \\
z_\downarrow & z_\uparrow^\ast
\end{array} \right) \rightarrow  V \left( \begin{array}{cc} z_\uparrow & -z_\downarrow^\ast \\
z_\downarrow & z_\uparrow^\ast
\end{array} \right)\,.
\end{align}
Section~\ref{sec:su2} describes how the SU(2) gauge symmetry is higgsed down to $\mathbb{Z}_2$ in states with fluctuating magnetic order, leading to the doping-induced $\mathbb{Z}_2$ fractionalized metal. The spectrum of spinless fermionic chargons in this state is the same as that of the electrons in the corresponding ordered magnetic state obtained in Hartree-Fock theory. 
We also present the low energy theory of bosonic $S=1/2$ spinons in the $\mathbb{Z}_2$ fractionalized metal, and estimate their spin gap. We noted above that the fractionalized state obtained in this manner from the spiral states of Fig.~\ref{fig:Fig2} have Ising-nematic order, while the state obtained from the canted state of Fig.~\ref{fig:canted} does not have any broken symmetry.

Section~\ref{sec:sb} presents an alternative theory of the fractionalized metal in terms of canonical Schwinger bosons $b_\sigma$ and a fermionic, spinless, chargon $f$ applied to the projected subspace of a $t$-$J$ model; so \cite{Lee89}
\begin{align}
c_\sigma = b_\sigma f^\dagger \quad, \quad b_\sigma^\dagger b_\sigma + f^\dagger f = 1
\label{cbf}
\end{align}
This parton decomposition only has a U(1) gauge symmetry under which $b_\sigma \rightarrow e^{i \theta} b_\sigma$, $f \rightarrow e^{i \theta} f$. We show in Section~\ref{sec:sb} that the U(1) is higgsed down to $\mathbb{Z}_2$, leading ultimately to a fractionalized state with the same underlying structure as that obtained from the canted state in the SU(2) gauge theory of the Hubbard model in Section~\ref{sec:su2}. Key to this conclusion is the presence of full square lattice symmetry and the absence of Ising-nematic order in the Schwinger boson spin liquid; furthermore, condensation of the  Schwinger bosons in this $\mathbb{Z}_2$ fractionalized state leads preferentially to the canted magnetic state in Fig.~\ref{fig:canted}.
The $f$ partons act as spinless fermions whose dispersion is shown in Fig.~\ref{fig: holon bands}, and feature a flat band at the Fermi level at $\nu=1/4$.

\section{Hartree-Fock theory}
\label{sec:hf}
\begin{figure}
    \centering
    \includegraphics[width=\linewidth]{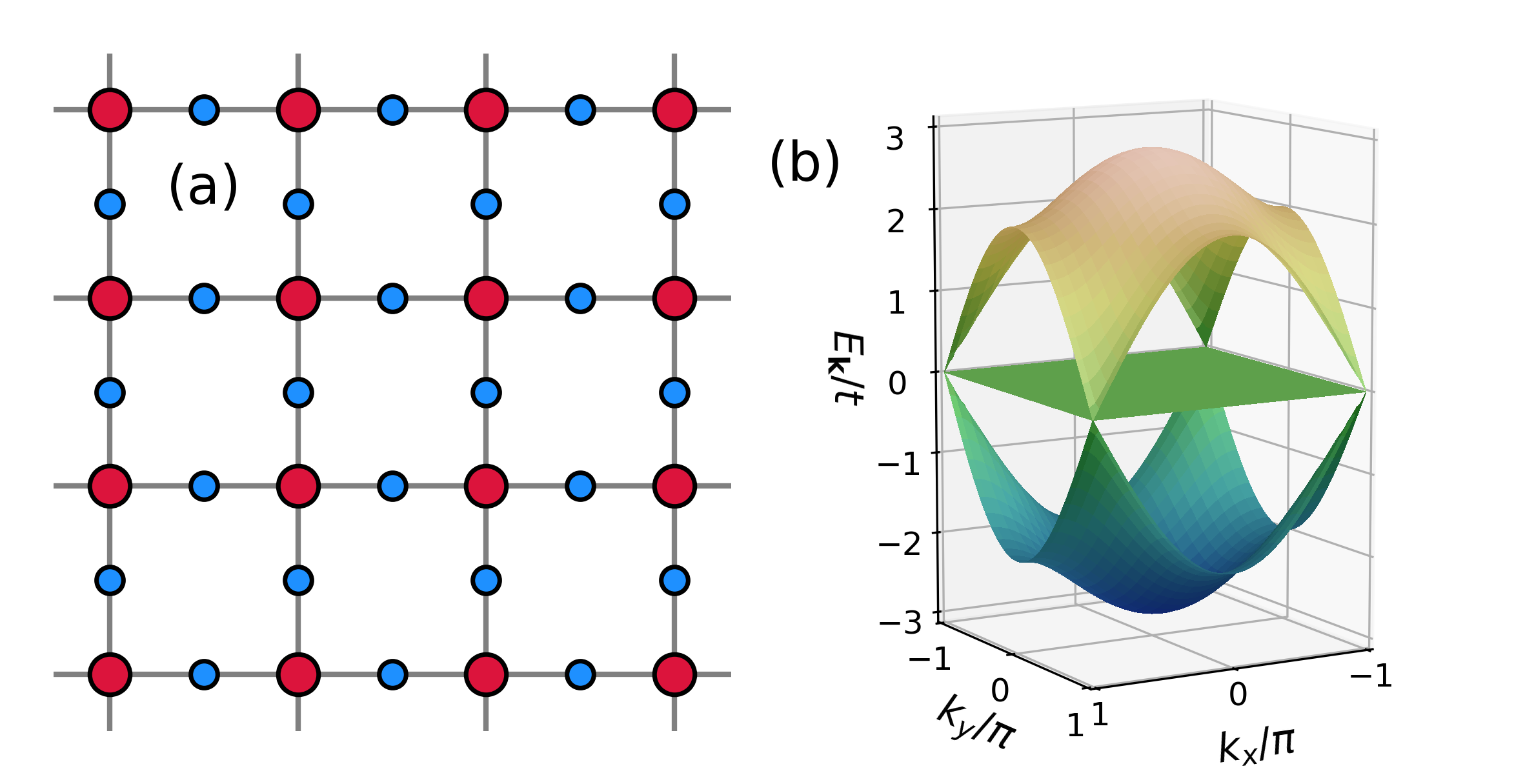}
    \caption{Panel (a): Lieb lattice. The sites of the $d$ sublattice are shown in red, while those of the $C_4$ related $p_x$ and $p_y$ sublattices are shown in blue. Panel (b):  Band structure at $U=0$. Note the presence of a flat band at $E_\bk=0$. }
    \label{fig: Lieb lattice}
\end{figure}
In this section, we apply the Hartree-Fock approximation to the Hubbard model on the Lieb lattice, which reads
\begin{equation}\label{eq: Hubbard model}
    H=-t\sum_{\langle j,j'\rangle,\sigma} c^\dagger_{j,\sigma}c_{j',\sigma} + U \sum_j n_{j,\up}n_{ij,\down}\,,
\end{equation}
where $c^\dagger_{j,\sigma}$ ($c_{j,\sigma}$) creates (annihilates) a fermion on Lieb lattice site $j$ with spin projection $\sigma$, the notation $\langle j,j' \rangle$ restricts the sum to those pairs of sites $j,j'$ that are nearest neighbors, $U>0$ is a contact repulsive interaction, and $n_{j,\sigma}=c^\dagger_{j,\sigma}c_{j,\sigma}\equiv\sum_\sigma n_{j,\sigma}$. In Fig.~\ref{fig: Lieb lattice}(a) we show a sketch of the Lieb lattice on which the $c_{j,\sigma}$ fermions live. This is composed of three sublattices, which we name $d$, $p_x$, and $p_y$, in analogy to the Emery model of cuprates. Note, however, that the lattice sites of our model contain $s$-orbitals, unlike the Emery model that actually has $d$ and $p$ orbitals on $d$ and $p_{x,y}$ sublattices, respectively. In the noninteracting limit ($U=0$) we can diagonalize~\eqref{eq: Hubbard model} exactly and obtain the band structure shown in Fig.~\ref{fig: Lieb lattice}(b). We get three bands: a perfectly flat band at zero energy and two particle-hole symmetric bands linearly touching the flat one at $\bk=(\pi,\pi)$.

Within the Hartree-Fock approximation, we mean-field decouple the Hubbard interaction in a spin-rotationally invariant way as~\cite{Gunnarsson1989,Scholle2023,Scholle2024}
\begin{equation}
    U n_{j,\up}n_{j,\down} \simeq -\dmu_j\, c^\dagger_j c_j + \vec{\Delta}_j\cdot c^\dagger_j \vec{\sigma} c_j +\frac{\dmu_j^2}{U} - \frac{|\vec{\Delta}_j|^2}{U}\,,
\end{equation}
where $\vec{\sigma}=(\sigma^1,\sigma^2,\sigma^3)$ are the Pauli matrices. The following self-consistency relations hold
\begin{subequations}\label{eq: MFeqs}
    \begin{align}
        & \dmu_j = -\frac{U}{2}\langle c^\dagger_j c_j\rangle\,,\\
        & \vec{\Delta}_j = \frac{U}{2}\langle c^\dagger_j \vec{\sigma} c_j\rangle\,.
    \end{align}
\end{subequations}
The expectation values are computed using the (quadratic) mean-field Hamiltonian
\begin{equation}
    H_\mathrm{MF} = -t\sum_{\langle j,j'\rangle} c^\dagger_{j}c_{j'} + \sum_j \left[-(\mu+\dmu_j)\, c^\dagger_j c_j + \vec{\Delta}_j\cdot c^\dagger_j \vec{\sigma} c_j \right]\,.
\end{equation}
In the above Hamiltonian we have added a chemical potential $\mu$ to fix the total filling $\nu=\frac{1}{2N_s}\sum_j \langle c^\dagger_j c_j\rangle$, where $N_s$ is the total number of sites. 

From this point on, we relabel the sites $j$ as $j=(\mathfrak{j},\alpha)$, where $\mathfrak{j}$ labels a Bravais vector, and $\alpha=d$, $p_x$, $p_y$ the three sites in the unit cell. The coordinates of site $j=(\mathfrak{j},\alpha)$ are expressed as $\bR_j=\bR_\mathfrak{j} +\mathbf{r}_\alpha \equiv x_\mathfrak{j} \hat{e}_x+y_\mathfrak{j} \hat{e}_y+\mathbf{r}_\alpha$, with $x_\mathfrak{j}$, $y_\mathfrak{j}$ integers, $\hat{e}_x=(1,0)$ and $\hat{e}_y=(0,1)$ are the elementary Bravais vectors, and $\mathbf{r}_d=(0,0)$, $\mathbf{r}_{p_x}=\hat{e}_x/2$, and $\mathbf{r}_{p_y}=\hat{e}_y/2$ are the sublattice shifts. Note that we have set the lattice spacing to $1/2$.

At this point, one can either solve equations~\eqref{eq: MFeqs} in real space on a finite-size system (see Refs.~\cite{Gunnarsson1989,Xu2011,Scholle2023} for such a computation on the square lattice and Ref.~\cite{Chiciak2018} for a three band Hubbard model on the Lieb lattice) or make an assumption on the spatial dependence of the parameters $\dmu_i$, $\vec{\Delta}_i$ and work in momentum space (see Refs.~\cite{Gouveia2015,Gouveia2016} for a similar computation on the Lieb lattice). Here, we take the second approach and make two different \textit{Ansätze} and retain the one that minimizes the Hartree-Fock free-energy:
\begin{itemize}
    \item We make no assumption on the form of $\dmu_j$, $\vec{\Delta}_j$ within the three-site unit cell, but we require that the spatial dependence of $\dmu_j$, $\vec{\Delta}_j$ can be obtained by replicating the unit cell with no modifications. We call this \textit{Ansatz} a $\bQ=0$ \textit{Ansatz}. In formulas, it reads
    $\dmu_{\mathfrak{j},\alpha}=\dmu_\alpha$, $\vec{\Delta}_{\mathfrak{j},\alpha}=\vec{\Delta}_\alpha$.
    \item We assume $\dmu_{\mathfrak{j},\alpha}=\dmu_\alpha$, and $\vec{\Delta}_j$ to take the form
    \begin{equation}
        \vec{\Delta}_{\mathfrak{j},\alpha}=A_{\alpha}\left[\cos(\bQ\cdot\bR_{\mathfrak{j},\alpha}+\varphi_\alpha)\, \hat{v}_1+\sin(\bQ\cdot\bR_{\mathfrak{j},\alpha}+\varphi_\alpha)\, \hat{v}_2\right]\,,
    \end{equation}
    where $A_\alpha$ and $\varphi_\alpha$ are some amplitudes and angles, respectively, that are allowed to vary only \textit{within} a unit cell, $\hat{v}_{1,2}$ are two mutually orthogonal unit vectors which can be arbitrarily chosen, and $\bQ=(Q_x,Q_y)$ is a wave-vector that minimizes the mean-field free energy. We call this \textit{Ansatz} a \textit{spiral Ansatz}. 
\end{itemize} 
Note that there is a particular class of states that can be described by \textit{both} \textit{Ans\"atze}. In fact, setting $\bQ=0$ in a spiral \textit{Ansatz}, is equivalent to requiring that a $\bQ=0$ state is coplanar, that is, all three spins in the unit cell lie within the same plane. 

\begin{figure}[t]
    \centering
    \includegraphics[width=1.0\textwidth]{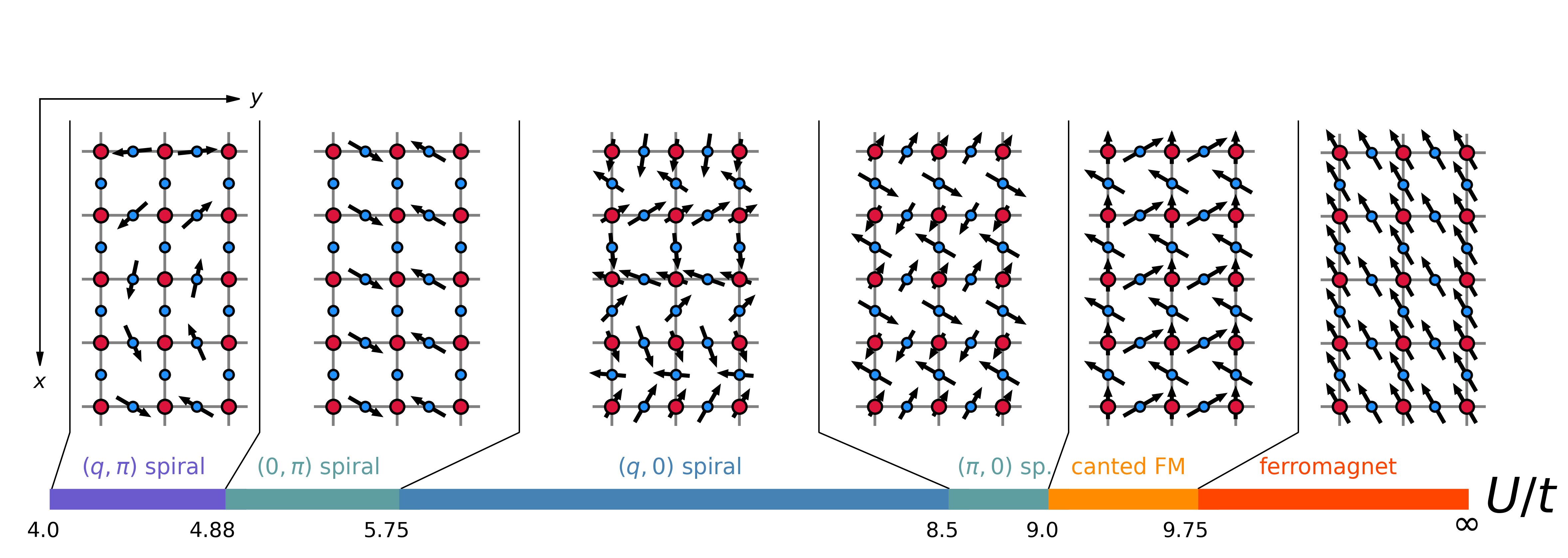}
    \caption{Phase diagram at quarter filling as a function of the Hubbard $U$. The insets above each phase represent a characteristic configuration of the spins in the corresponding phase. Note that in the insets, for representation purposes, the $x$-axis is oriented vertically and the $y$-axis horizontally.}
    \label{fig:Fig2}
\end{figure}

We perform a scan in the interaction parameter $U$ at quarter filling $\nu=1/4$ and fixed (low) temperature $T=0.01t$ and for two system sizes: 20$\times$20 and 40$\times$40 unit cells. In the following we are showing results for the 40$\times$40 system, but the phase diagram for the 20$\times$20 one is qualitatively identical.

In Fig.~\ref{fig:Fig2} we show the Hartree-Fock phase diagram at quarter filling as a function of the interaction strength $U$ in units of the hopping parameter $t$. For $U<4t$ we find a paramagnetic solution, that is, where $\vec{\Delta}_j=0$. We find magnetic order to set in at $U\approx 4t$, where a spiral phase with $\bQ=(q,\pi)$ (or symmetry related) emerges. Upon increasing $U$, $q$ gradually decreases, becoming zero at $U\approx 4.88\,t$. In the regime $4t\lesssim U\lesssim 5.75t$ we find a nonvanishing magnetization only on $p_y$ sites (or $p_x$ sites, if $\bQ=(\pi,q)$). For $5.75t\lesssim U\lesssim 9t$ the magnetization on $p_x$ and $d$  deviates from zero and so does $q$, first decreasing from $q=\pi$ at $U=5.75$ to $q\approx 0.67 \pi$ at $U=6t$ and then increasing again to $q=\pi$ at $U\approx 8.5 t$. For $8.5t\lesssim U\lesssim 9t$, $q$ remains locked to $\pi$. All phases described so far spontaneously break the lattice point group symmetry $C_{4v}$ and all three amplitudes (if finite) $A_d$, $A_{p_x}$, $A_{p_y}$ take different values. Similarly, the local densities are different on each one of the sublattices. We also find $\varphi_{p_x}=\varphi_d+\pi$ and $\varphi_{p_y}=\varphi_d$. Typical spin patterns of the spiral phases are shown in the four leftmost insets of Fig.~\ref{fig:Fig2}.

At $U\approx 9t$ we find a first order transition to a $\bQ=0$ state to occur. The phase found immediately after such transition happens to be the canted ferromagnetic state displayed in Fig.~\ref{fig:canted}, characterized by equal spin amplitudes on $p_x$ and $p_y$ sites, that is, $A_{p_x}=A_{p_y}\neq A_d$ but opposite phases, $\varphi_{p_x}=\varphi_d+\theta$ and $\varphi_{p_y}=\varphi_d-\theta$. We also find equal densities on $p_x$ and $p_y$ sites. $\theta$ is a continuously varying parameter, ranging from $\theta \approx 0.24\pi$ at $U\approx 9t$ to $\theta \approx 0.16\pi$ at $U\approx 9.75t$, beyond which it discontinuously jumps to zero. For $U\gtrsim 9.75$ we find a fully polarized ferromagnet with saturated spin amplitudes $A_\alpha=\frac{U}{4}$ for all $\alpha$. Sketches of the spin patterns in the canted ferromagnetic and ferromagnetic phases are shown in the two rightmost insets of Fig.~\ref{fig:Fig2}.

\subsection{Spectral properties of the canted ferromagnetic phase}
\label{subsec: spectrum canted FM}
\begin{figure}[t]
    \centering
    \includegraphics[width=1.0\linewidth]{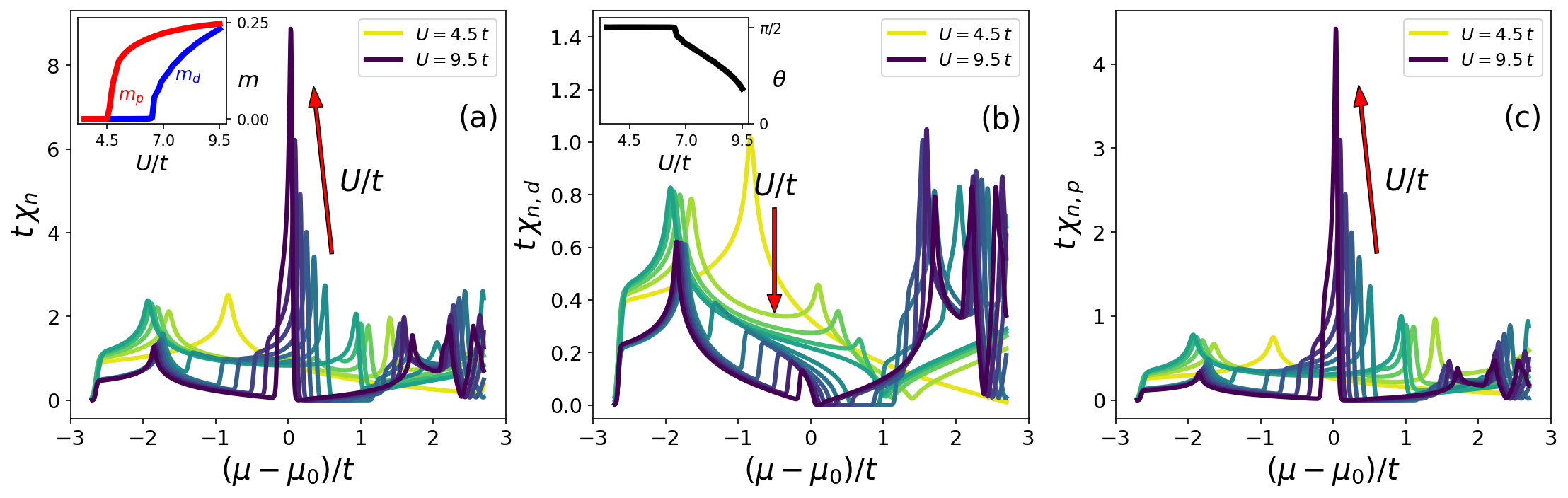}
    \caption{Total compressibility $\chi_n$ (panel (a)), $d$-site compressibility $\chi_{n,d}$ (panel (b)), and $p$-site compressibility (panel (c)) as functions of the chemical potential $\mu$ for several values of the Hubbard $U$. Here, $\mu_0$ indicates the ($U$-dependent) value of the chemical potential that enforces quarter filling. Inset of panel (a): magnetizations on $d$- and $p$-sites as functions of $U$. Inset of panel (b): angle $\theta$ (see text) as a function of $U$.}
    \label{fig:Fig3}
\end{figure}
\begin{figure}[t]
    \centering
    \includegraphics[width=1.0\linewidth]{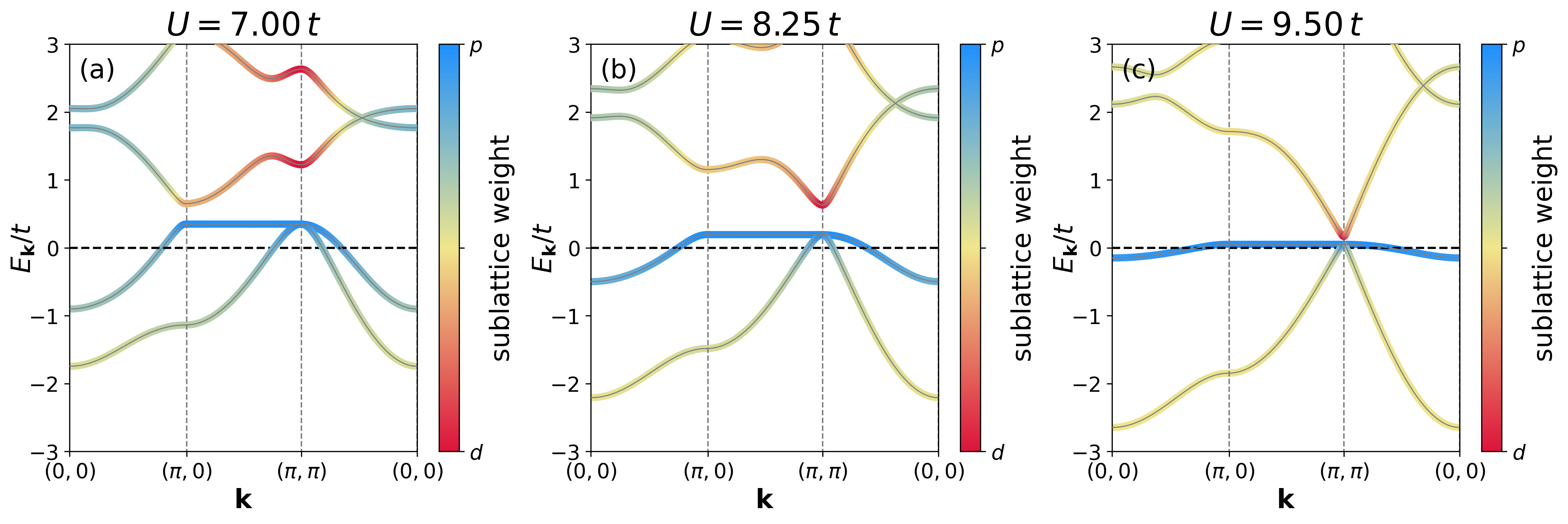}
    \caption{Band structures in the canted ferromagnetic phase for different values of the Hubbard $U$, indicated on top of each panel. The color of the bands in each panel indicates the sublattice weight at every $\bk$ point. The parameters of the calculation are shown in the insets of Fig.~\ref{fig:Fig3}.}
    \label{fig: bandstructures}
\end{figure}
It is instructive to analyze the spectral properties of the magnetic phases found within Hartree-Fock at quarter-filling. As a proxy of the spectral function, we define the compressibility $\chi_n={\partial n}/{\partial \mu}$. Furthermore,we define the \textit{sublattice-resolved} compressibility $\chi_{n,\alpha}={\partial n_\alpha}/{\partial \mu}$, with $\alpha=d,\,p_x$ or $p_y$. These two quantities were measured in the cold atomic simulator of Ref.~\cite{Lebrat24}. We compute the (sublattice-resolved) compressibility at fixed value of the interaction strength by keeping the self-consistently determined parameter $\vec{\Delta}_{\mathfrak{j},\alpha}$ fixed to their Hartree Fock values and vary only the chemical potential $\mu$ to evaluate the total density $n(\mu)$ as well as the sublattice-resolved densities $n_\alpha(\mu)$. We then numerically differentiate these functions to obtain the compressibilities. In the following, we present results obtained by forcing a $\bQ=0$ solution. 

Imposing such solution, we always find the canted ferromagnetic state where $A_{p_x}=A_{p_y}\equiv A_p$ and $\varphi_{p_x}=\varphi_d+\theta$, $\varphi_{p_y}=\varphi_d-\theta$.
In the inset of panel (a) of Fig.~\ref{fig:Fig3} we show the magnetizations on $d$- ($m_d$) and $p$-sites ($m_p$) as functions of $U$, whereas 
in the inset of panel (b) of Fig.~\ref{fig:Fig3} $\theta$ as a function of $U$. 
We start finding a magnetic state at $U=4.5t$, where $A_{d}=0$ and $\theta=\pi/2$. At $U=6.5t$ the magnetization on the $d$ sites starts to be nonzero and $\theta$ deviates from $\pi/2$. Eventually, at $U=9.75t$, $m_d$ and $m_p$ reach their saturation value $1/4$ and a for larger $U$ (not shown) we get a fully polarized ferromagnetic state. 

Upon increasing the Hubbard $U$ we observe that the total compressibility (panel (a) of Fig.~\ref{fig:Fig3}) develops a sharp peak around $\mu=0$, signaling the flattening of the band at the Fermi level. When the system transitions to the ferromagnetic state, such a band becomes completely flat and the compressibility exhibits a Dirac delta-like structure at $\mu=0$. Inspecting the site-resolved compressibilities $\chi_{n,d}$ and $\chi_{n,p}$ (panels (b) and (c) of Fig.~\ref{fig:Fig3}), we note that the zero-energy peak in the total compressibility originates from the compressibility on $p$ sites. Furthermore, we note that the $d$-site compressibility develops a dip (and eventually vanishes) around $\mu=\mu_0$ (with $\mu_0$ being the value of the chemical potential at quarter filling) as interactions are increased. We note that the relation $\chi_n=\chi_{n,d}+2\chi_{n,p}$ holds.

To further investigate the origin of the peak in compressibility, in Fig.~\ref{fig: bandstructures} we plot the band structure of the canted ferromagnetic phase along the Brillouin zone path $(0,0)$-$(\pi,0)$-$(\pi,\pi)$-$(0,0)$ for three different values of $U/t$. For $U=7t$, we observe a dispersive band around the Fermi level that is flat along the $(\pi,0)$-$(\pi,\pi)$ path. Upon increasing the Hubbard $U$ the band level flattens everywhere in the Brillouin zone and gets an increasingly larger weight on $p$-sites. In the ferromagnetic phase (not shown) the band becomes exactly flat and its weight on the $d$ sublattice becomes exactly zero. This can be simply understood by the fact that the bandstructure in the ferromagnetic phase is nothing but the noninteracting band structure with a momentum independent spin split. It is then natural that close to the ferromagnetic phase, that is, when the canting angle $\theta$ is small, the flat band acquires a small dispersion, growing with increasing $\theta$.

\subsection{Spectral properties in the spiral phases}
\begin{figure}
    \centering
    \includegraphics[width=1\linewidth]{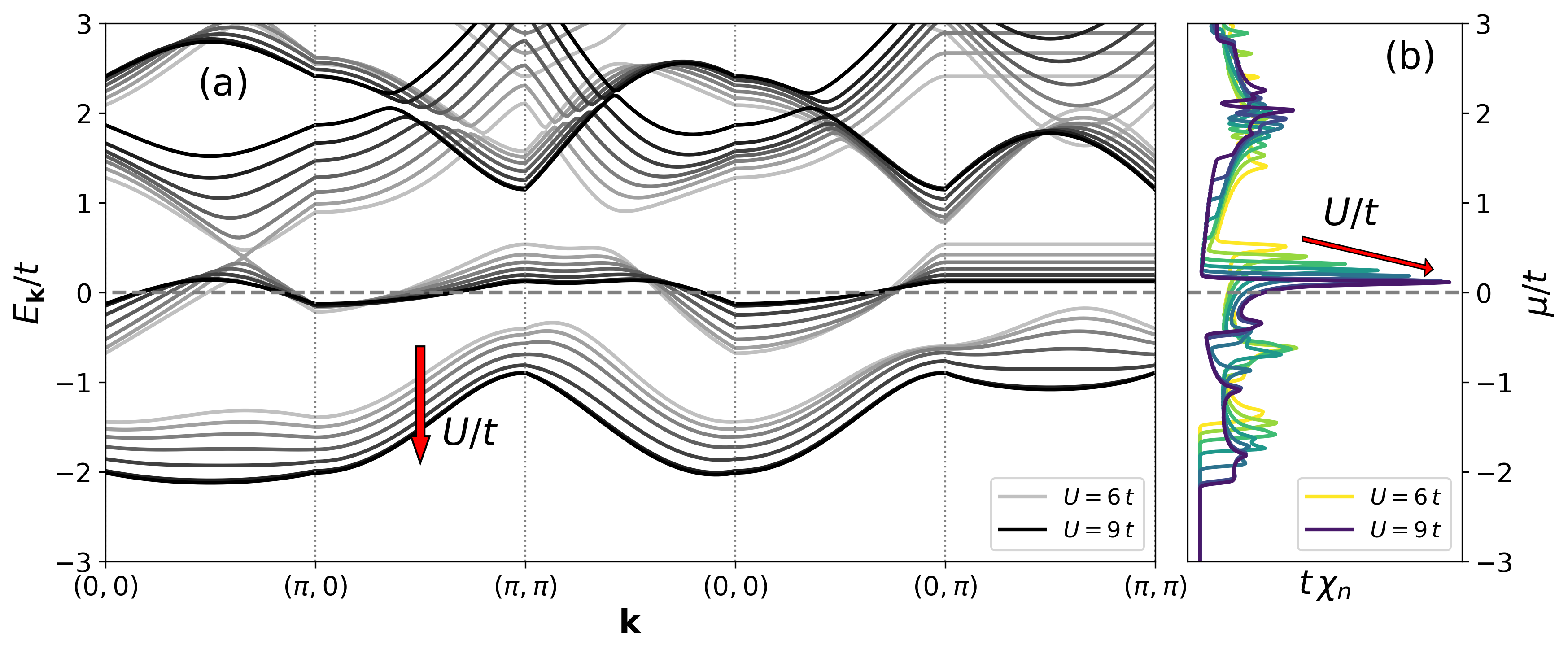}
    \caption{Band structures (panel (a)) and compressibilities (panel (b)) for $(q,0)$ and $(\pi,0)$ spiral phases for several values of the interaction strength, ranging from $6t$ to $9t$. }
    \label{fig: spectrum spiral}
\end{figure}
We here briefly discuss the behavior of spectral properties in the spiral phases. In Fig.~\ref{fig: spectrum spiral}, we show band structures (panel (a)) and compressibilities (panel (b)) for different values of the interaction $U$. We observe that larger values of $U$ give rise to a flatter band around the Fermi level, seen as a peak in the compressibility. However, at least within the interaction interval considered here, such band remains rather dispersing. 

\section{SU(2) gauge theory of fermionic chargons and bosonic spinons}
\label{sec:su2}

A possible way to study quantum fluctuations on top of the Hatree-Fock state is by resorting to an SU(2) gauge theory~\cite{sdw09,DCSS15b,CSS17,Scheurer:2017jcp,Sachdev:2018ddg,MSSS18,SSST19,ScheurerShen,Bonetti22}. We start with Eq.~\eqref{czpsi} , where we \textit{fractionalize} the Hubbard model fermions into bosonic spinons $z_{\sigma}$, representing fluctuations of the spin reference frame~\cite{SS80,Schulz90,Dupuis02,Dupuis04} and spinless fermionic chargons $\psi_\pm$, carrying the charge of the original fermions.

We perform a Hubbard-Stratonovich transformation on the imaginary time action of the Hubbard model in the spin and charge channels while preserving spin-rotational invariance (see Ref.~\cite{Dupuis04} for details), we get
\begin{equation}
    \begin{split}
        \mathcal{S}[c,\bar{c},\vec{S},\rho] = \int_0^\beta\!d\tau\,\bigg[\sum_j \bar{c}_j(\partial_\tau-\mu)c_j-t\sum_{\langle j,j'\rangle}\bar{c}_j c_{j'}-\sum_j \left(i\rho_j\, \bar{c}_j c_j+\vec{S}_j\cdot\bar{c}_j\vec{\sigma}c_j\right)+\frac{1}{U}\sum_j\left(\rho_j^2+|\vec{S}_j|^2\right)\bigg]\,,
    \end{split}
\end{equation}
with $\beta=1/T$ the inverse temperature. Applying decomposition~\eqref{czpsi} to the above action, we obtain
\begin{equation}\label{eq: chargon spinon HS action}
    \begin{split}
        \mathcal{S}[\psi,\bar{\psi},R,\vec{H},\rho] = \int_0^\beta\!d\tau\,\bigg[&\sum_j \bar{\psi}_j(\partial_\tau+R^\dagger_j\partial_\tau R_j-\mu)\psi_j-t\sum_{\langle j,j'\rangle}\bar{\psi}_j R^\dagger_j R_{j'
}\psi_{j'}\\
-&\sum_j \left(i\rho_j\, \bar{\psi}_j \psi_j+\vec{H}_j\cdot\bar{\psi}_j\vec{\sigma}\psi_j\right)+\sum_j\left(\frac{\rho_j^2}{U}+\frac{|\vec{H}_j|^2}{U}\right)\bigg]\,,
    \end{split}
\end{equation}
where
\begin{align}
    R_j=\left(\begin{matrix} z_{j,\up} & - z_{j,\down}^* \\ z_{j,\down} & z_{j,\up}^*\end{matrix}\right)\,,
\end{align}
is an SU(2) matrix, implying $|z_{j,\up}|^2+|z_{j,\down}|^2=1$, and 
\begin{align}
    \vec{H}_j\cdot\vec{\sigma}\equiv R^\dagger_j \vec{\sigma} R_j\cdot\vec{S}_j\,,
\end{align} is a spinless Higgs field. In Tab.~\ref{tab:psiRH transformations} we list the transformation properties of the fields in the above action as well as of the Hubbard model fermions $c_j$ and spin field $\vec{S}_j$ under global spin transformations, SU(2) gauge transformations, and global U(1) charge transformations. As discussed in Sec.~\ref{sec:intro}, decomposition~\eqref{czpsi} brings in an SU(2) gauge redundancy, which will lead to the emergence of an SU(2) massless gauge field $A_\mu^{a=1,2,3}$~\cite{sdw09}. 

\begin{table}[t!]
    \centering
    \begin{tabular}{|c||c|c|c|}
         \hline & SU(2)$_s$ & SU(2)$_g$ & U(1)$_c$\\\hline \hline
        $c_j$ & $V c_j$ & $c_j$ & $e^{i\theta}c_j$\\\hline
        $\vec{S}_j$ & $\mathcal{V} \vec{S}_j$ & $\vec{S}_j$ & $\vec{S}_j$\\\hline
        $\psi_j$ & $\psi_j$ & $U_j\psi_j$ & $e^{i\theta}\psi_j$\\\hline
        $R_j$ & $VR_j$ & $R_j U_j^\dagger$ & $R_j$\\\hline
        $\vec{H}_j$ & $\vec{H}_j$ & $\mathcal{U}_j\vec{H}_j$ & $\vec{H}_j$\\\hline
        $\rho_j$ & $\rho_j$ & $\rho_j$& $\rho_j$\\\hline
    \end{tabular}
    \caption{Transformation properties of the Hubbard model fermions $c_j$ and spin field $\vec{S}_j$ and of the fields $\psi=(\psi_+,\psi_-)$, $R_j$, $\vec{H}_j$, and $\rho_j$ under global spin rotations (SU(2)$_s$), SU(2) gauge transformations (SU(2)$_g$), and global U(1) charge transformations (U(1)$_c$). Here $V$ and $U_j$ are SU(2) matrices, and $\mathcal{V}_{ab}=\frac{1}{2}\left[\sigma^a V^\dagger \sigma^b V\right]$ and $\mathcal{U}_{j,ab}=\frac{1}{2}\left[\sigma^a U^\dagger_j \sigma^b U_j\right]$ are the adjoint representations of $V$ and $U_j$, respectively.}
    \label{tab:psiRH transformations}
\end{table}

The condensation of $\vec{H}_j$ will \textit{higgs down} the gauge group to U(1) or even $\mathbb{Z}_2$, depending on the specific form of $\vec{H}_j$, by gapping out two (for U(1)) or all three (for $\mathbb{Z}_2$) components of $A_\mu^a$, leaving behind a state with U(1) or $\mathbb{Z}_2$ topological order. The `vison' or $m$ particle of the $\mathbb{Z}_2$ fractionalized state is realized by a vortex configuration of the Higgs field \cite{SSST19}.
Note that the U(1) phase might be eventually unstable against the formation of valence bond solid (VBS) order due to the proliferation of monopoles~\cite{ReadSachdev1989}.
The additional condensation of $R_j$ in a phase with $\avH{j}\neq 0$ fully confines the U(1) or $\mathbb{Z}_2$ gauge field and spontaneously breaks the SU(2) global spin symmetry, giving a state with long range magnetic order. The topologically ordered ($\avH{j}\neq 0$, $\langle R_j\rangle= 0$) phase is separated from a trivial paramagnetic phase ($\avH{j} = 0$, $\langle R_j\rangle\neq 0$) by a deconfined quantum critical point (DQCP) where $\avH{j}=0$ and $\langle R_j\rangle= 0$.

To make further progress, we treat the charge field $\rho$ in action~\eqref{eq: chargon spinon HS action} within a mean-field approximation, that is, we perform the replacement $i\rho_j\to \dmu_j\equiv -\frac{U}{2}\langle \bar{\psi}_j\psi_j\rangle$. Similarly, we can replace the Higgs field with its mean-field expectation value $\langle \vec{H}_j \rangle=\frac{U}{2}\langle\bar{\psi}_j\vec{\sigma}\psi_j\rangle$. Moreover, following Ref.~\cite{CSS17}, we decouple in a mean-field fashion all those terms containing two $R$ fields and two $\psi_\pm$ fields, getting
\begin{subequations}\label{eq: chargon spinon MF action}
    \begin{align}
        \mathcal{S}_\mathrm{eff}[\psi,\bar{\psi},R] =& \mathcal{S}_\psi[\psi,\bar{\psi}] + \mathcal{S}_z[z,z^*]\,,\\
        \mathcal{S}_\psi[\psi,\bar{\psi}]=&\int_0^\beta\!d\tau\,\bigg\{\sum_j \bar{\psi}_j[\partial_\tau-(\mu+\dmu_j)-\langle\vec{H}_j\rangle\cdot\vec{\sigma}]\psi_j-t\sum_{\langle j,j'\rangle}\bar{\psi}_j\, T^R_{jj'}\,\psi_{j'}\bigg\}\,,\label{eq: MF chargon action}\\
        \mathcal{S}_z[z,z^*]=&\int_0^\beta\!d\tau\,\bigg\{\sum_j \left[(\chi_{jj}^{++}-\chi_{jj}^{--})z_j^*\partial_\tau z_j - \chi_{jj}^{-+} z_j\,\varepsilon\,\partial_\tau z_j+ \chi_{jj}^{+-} z^*_j\,\varepsilon\,\partial_\tau z^*_j-\bar{\lambda}_j z^*_j z_j\right]\label{eq: MF spinon action}\\
        &\hskip 5mm -t\sum_{\langle j,j'\rangle}\left[(\chi_{jj'}^{++}+\chi_{j'j}^{--})z_j^* z_{j'}-(\chi_{jj'}^{-+}-\chi_{j'j}^{-+})z_j\,\varepsilon\,z_{j'}+(\chi_{jj'}^{+-}-\chi_{j'j}^{+-})z^*_j\,\varepsilon\,z^*_{j'}\right]\bigg\}\nonumber\,,        
    \end{align}
\end{subequations}
where $\varepsilon=\left(\begin{smallmatrix}0&-1\\1&0\end{smallmatrix}\right)$, $\chi_{jj'}^{ss'}=\langle \bar{\psi}_{j,s}\psi_{j',s'}\rangle$, $T^R_{jj'}=\langle R^\dagger_j R_{j'}\rangle$ and we have re-expressed $R_j$ as $\left(\begin{smallmatrix} z_{j,\up} & - z_{j,\down}^* \\ z_{j,\down} & z_{j,\up}^*\end{smallmatrix}\right)$. We have also added a Lagrange multiplier field $\lambda_j$ to enforce the constraint $|z_{j,\up}|^2+|z_{j,\down}|^2=1$, and replaced it with its expectation value $\bar{\lambda}_j=-i\langle \lambda_j\rangle$. Note that, due to the static nature of the mean-field approximation, $\langle R^\dagger_j\partial_\tau R_j\rangle=0$. Assuming $T_{jj'}^R=Z_{j-j'}\mathbbm{1}$~\cite{CSS17,Scheurer:2017jcp}, where $0< Z_{j-j'}< 1$ is a dimensionless number renormalizing the chargon hopping parameters, one can translate the Hartree-Fock results to the action above with the replacements $t\to t_\mathrm{eff}=Z_{\langle j,j'\rangle} t<t$, $c_j\to\psi_j$, $\vec{\Delta}_j\to\langle\vec{H}_j\rangle$, while $\dmu_j$ remains the same. 
The main difference between the Hartree-Fock method and the SU(2) gauge theory is that the latter includes quantum fluctuations through the spinons and can therefore host nontrivial phases with no magnetic order. 

We now analyze the form that the effective action $\mathcal{S}_\mathrm{eff}$ takes in each one of the phases found within Hartree-Fock approximation for $\avH{j}$.

\subsection{Canted Ferromagnetic Higgs phase}
Within our gauge choice, the canted ferromagnetic phase is characterized by
\begin{equation}
    \avH{j}=\avH{\mathfrak{j},\alpha}=
    \begin{cases}
        H_d(0,0,1) &\alpha=d\,, \\
        H_p(\sin\theta,0,\cos\theta) \quad &\alpha=p_x\,,\\         
        H_p(-\sin\theta,0,\cos\theta)\quad &\alpha=p_y\,,
    \end{cases}
\end{equation}
with $H_\alpha>0$. It is convenient to perform a gauge transformation $\psi_j\to V_j \psi_j$, with $V_{\mathfrak{j},\alpha}=e^{-i\frac{\phi_\alpha}{2}\sigma^2}$, $\phi_\alpha=\{0,\theta,-\theta\}$, that aligns the Higgs field along the $(0,0,1)$ direction on all sites at the expense of creating a hopping term that is non-diagonal in the SU(2) gauge indices $s,s'$. In momentum space, the chargon action reads
\begin{equation}\label{eq: canted FM fermion action}
    \begin{split}
        \mathcal{S}_\psi[\psi,\bar{\psi}]=-T\sum_{\omega_n}\int_\bk\,\Psi^\dagger_{k} \left[i\omega_n\mathbbm{1}_6-\left(
        \begin{array}{c|c}
            A^\theta_\bk - H & B_\bk^\theta \\ \hline
            (B_\bk^\theta)^\dagger & A^\theta_\bk+H
        \end{array}
        \right)\right]\Psi_{k}\,,
    \end{split}
\end{equation}
where $k=(\bk,\omega_n)$, $\omega_n=(2n+1)\pi T$ indicates the fermionic Matsubara frequencies, and 
\begin{subequations}
    \begin{align}
        \Psi_{k}=&(\psi_{k,d,+}\quad\psi_{k,p_x,+}\quad\psi_{k,p_y,+}\quad\psi_{k,d,-}\quad\psi_{k,p_x,-}\quad\psi_{k,p_y,-})^T\,,\\
        A^\theta_\bk=&\left(
        \begin{array}{ccc}
            -(\mu+\dmu_d) & -2t_\mathrm{eff}\cos\left(\frac{k_x}{2}\right)\cos\left(\frac{\theta}{2}\right) & -2t_\mathrm{eff}\cos\left(\frac{k_y}{2}\right)\cos\left(\frac{\theta}{2}\right)\\
            -2t_\mathrm{eff}\cos\left(\frac{k_x}{2}\right)\cos\left(\frac{\theta}{2}\right) & -(\mu+\dmu_p) & 0 \\
            -2t_\mathrm{eff}\cos\left(\frac{k_y}{2}\right)\cos\left(\frac{\theta}{2}\right)& 0 & -(\mu+\dmu_p) \\
        \end{array}
        \right)\,,\\
        B^\theta_\bk=&\left(
        \begin{array}{ccc}
            0 & -2t_\mathrm{eff}\cos\left(\frac{k_x}{2}\right)\sin\left(\frac{\theta}{2}\right) & +2t_\mathrm{eff}\cos\left(\frac{k_y}{2}\right)\sin\left(\frac{\theta}{2}\right)\\
            +2t_\mathrm{eff}\cos\left(\frac{k_x}{2}\right)\sin\left(\frac{\theta}{2}\right) & 0 & 0 \\
            -2t_\mathrm{eff}\cos\left(\frac{k_y}{2}\right)\sin\left(\frac{\theta}{2}\right)& 0 & 0 \\
        \end{array}
        \right)\,, \label{eq: B matrix definition}\\
        H=&\mathrm{diag}(H_d,H_p,H_p)\,. 
    \end{align}
\end{subequations}

The present choice of $\avH{j}$ retains translational invariance in $\mathcal{S}_\psi$, implying that $\chi_{jj'}^{ss'}=\chi_{\alpha\alpha'}^{ss'}(\bR_{\mathfrak{j},\alpha}-\bR_{\mathfrak{j}',\alpha'})$. Owing to this symmetry, we define $\chi_\alpha^{ss'}=\chi^{ss'}_{\alpha\alpha}(0)=[\chi^{s's}_{\alpha\alpha}(0)]^*$ and $\chi^{ss'}_{\mu,\pm}=\chi_{d,p_\mu}^{ss'}(\mp\hat{e}_\mu/2)=[\chi_{p_\mu,d}^{s's}(\pm\hat{e}_\mu/2)]^*$, 
with $\mu=x,y$. The system's $C_{2z}$ symmetry $\psi_{j,s}\to \psi_{-j,s}$ further enforces $\chi^{ss'}_{\mu,+}=\chi^{ss'}_{\mu,-}\equiv\chi^{ss'}_{\mu}$. $\mathcal{S}_\psi$ is also symmetric under the antiunitary transformation $\Theta = \mathbbm{1}\mathcal{K}$ (see Tab.~\ref{tab: transformations Z2 QSL}), with $\mathcal{K}$ denoting complex conjugation, implying that $\chi^{ss'}_X\in\mathbb{R}$, with $X=d,p_x,p_y,x,y$. Additionally, the chargon action $\mathcal{S}_\psi$ is invariant under the action of the symmetry $\psi_{\mathfrak{j},\alpha,s}\to S_{\alpha\alpha'}[e^{i\frac{\pi}{2}\sigma^3}]_{ss'}\psi_{C_{4z}\mathfrak{j},\alpha',s'}$ (see Tab.~\ref{tab: transformations Z2 QSL}), implying $\chi_d=\mathrm{diag}(\chi_d^{++},\chi_d^{--})$, $\chi^{ss'}_{p_x}=ss'\chi^{ss'}_{p_y}$, $\chi^{ss'}_{x}=ss'\chi^{ss'}_{y}$. We further notice that $\chi_\alpha^{ss'}=\chi^{(0)}_\alpha+\vec{\chi}_\alpha\cdot\vec{\sigma}$, where $\vec{\chi}_\alpha\propto \avH{\alpha}$. Because we chose a gauge in which $\avH{j}$ is parallel to $(0,0,1)$ on every site, all $\chi_\alpha^{ss'}$ will be diagonal. 
Combining all the symmetries, we find the spinon action to take the following form
\begin{equation}\label{eq: canted FM boson action}
    \begin{split}
        \mathcal{S}_z[z,z^*]=-T\sum_{\Omega_n}\int_\bk\,Z^\dagger_{k} \left[i\chi_\tau\Omega_n-\left(
        \begin{array}{c|c}
            P_\bk & Q_\bk \\ \hline
            Q_\bk^\dagger &P_{-\bk}^T
        \end{array}
        \right)\right]Z_{k}\,,
    \end{split}
\end{equation}
where $k=(\bk,\Omega_n)$, $\Omega_n=2n\pi T$ indicates the bosonic Matsubara frequencies and
\begin{subequations}
    \begin{align}
    Z_{k}=&(z_{k,d,\up}\quad z_{k,p_x,\up}\quad  z_{k,p_y,\up} \quad z^*_{-k,d,\down}\quad z^*_{-k,p_x,\down} \quad z^*_{-k,p_y,\down})^T\,,\\
        P_\bk=&\left(
        \begin{array}{ccc}
            -\bar{\lambda}_d & 2P\cos\left(\frac{k_x}{2}\right) & 2P\cos\left(\frac{k_y}{2}\right)\\
            2P\cos\left(\frac{k_x}{2}\right) & -\bar{\lambda}_p & 0 \\
            2P\cos\left(\frac{k_y}{2}\right)& 0 & -\bar{\lambda}_p \\
        \end{array}
        \right)\,,\label{eq: Pk bosonic action canted FM Higgs}\\
        Q_\bk=&\left(
        \begin{array}{ccc}
            0 & +2Q\cos\left(\frac{k_x}{2}\right) & -2Q\cos\left(\frac{k_y}{2}\right)\\
            -2Q\cos\left(\frac{k_x}{2}\right) & 0 & 0 \\
            +2Q\cos\left(\frac{k_y}{2}\right)& 0 & 0 \\
        \end{array}
        \right)\,,\label{eq: Qk bosons SU(2)}\\
        \chi_\tau=&\mathrm{diag}(\chi_{d,\tau},\chi_{p,\tau},\chi_{p,\tau},-\chi_{d,\tau},-\chi_{p,\tau},-\chi_{p,\tau})\,,
    \end{align}
\end{subequations}
where $P=-t(\chi_x^{++}+\chi_x^{--})$, $Q=-2t\chi_x^{+-}$, and $\chi_{\alpha,\tau}=\chi_\alpha^{++}-\chi_\alpha^{--}$. We observe that, when $\avH{j}$ is calculated within Hartree-Fock theory, than $Q/P=\tan(\theta/2)$.

Action~\eqref{eq: canted FM boson action} describes a $\mathbb{Z}_2$ spin liquid that preserves all original space group symmetries. While it is obvious to see that translations are preserved by action~\eqref{eq: canted FM boson action}, $C_{4z}$ rotations and $\sigma_d$ diagonal mirror operations act projectively, as listed in Tab.~\ref{tab: transformations Z2 QSL}.
\begin{table}[t!]
    \centering
    \begin{tabular}{|c||c|c|c|c|}
        \hline & $T$ & $C_{4z}$ & $\sigma_d$ & $\Theta$\\\hline
        $\bR_{\mathfrak{j},\alpha}=(x_\mathfrak{j},y_\mathfrak{j})+\mathbf{r}_\alpha$ & $\bR_j+\mathbf{a}$ & $(-y_\mathfrak{j},x_\mathfrak{j})+S_{\alpha\alpha'}\mathbf{r}_{\alpha'}$ & $(y_\mathfrak{j},x_\mathfrak{j})+S_{\alpha\alpha'}\mathbf{r}_{\alpha'}$ & $\bR_{\mathfrak{j},\alpha}$\\ \hline
        $c_{\mathfrak{j},\alpha}$ & $c_{\mathfrak{j}+\mathbf{a},\alpha}$ & $S_{\alpha\alpha'}\,c_{C_{4z}\mathfrak{j},\alpha'}$& $S_{\alpha\alpha'}\,c_{\sigma_d\mathfrak{j},\alpha'}$ & $i\sigma^2 c_{\mathfrak{j},\alpha}\,\mathcal{K}$ \\ \hline
        $\psi_{\mathfrak{j},\alpha}$ & $\psi_{\mathfrak{j}+\mathbf{a},\alpha}$ & $S_{\alpha\alpha'}\,e^{i\frac{\pi}{2}\sigma^3}\,\psi_{C_{4z}\mathfrak{j},\alpha'}$& $S_{\alpha\alpha'}\,e^{i\frac{\pi}{2}\sigma^3}\,\psi_{\sigma_d\mathfrak{j},\alpha'}$ & $\psi_{\mathfrak{j},\alpha}\,\mathcal{K}$\\ \hline
        $R_{\mathfrak{j},\alpha}$ & $R_{\mathfrak{j}+\mathbf{a},\alpha}$ & $S_{\alpha\alpha'}\,R_{C_{4z}\mathfrak{j},\alpha'}\,e^{-i\frac{\pi}{2}\sigma^3}$& $S_{\alpha\alpha'}\,R_{\sigma_{d}\mathfrak{j},\alpha'}\,e^{-i\frac{\pi}{2}\sigma^3}$ & $i\sigma^2 R_{\mathfrak{j},\alpha}\mathcal{K}$ \\ \hline
        $z_{\mathfrak{j},\alpha}$ & $z_{\mathfrak{j}+\mathbf{a},\alpha}$ & $S_{\alpha\alpha'}\,e^{-i\frac{\pi}{2}}\,z_{C_{4z}\mathfrak{j},\alpha'}$& $S_{\alpha\alpha'}\,e^{-i\frac{\pi}{2}}\,z_{\sigma_d\mathfrak{j},\alpha'}$ & $i\sigma^2 z_{\mathfrak{j},\alpha}\mathcal{K}$\\ \hline
    \end{tabular}
    \caption{Transformation properties of the $\mathbb{Z}_2$ spin liquid of Eqs.~\eqref{eq: canted FM fermion action} and \eqref{eq: canted FM boson action} under translations ($T$), 90 degree rotations ($C_{4z}$), diagonal mirror symmetries ($\sigma_d$) and time reversal $\Theta$. In the first row we list the transfromation properties of the lattice coordinates, in the second one those of the original Hubbard model fermions $c_j$,  in the third one those of the fermionic chargons $\psi_j$, in the fourth one those of the bosonic spinons $R_j$. Note that the fifth and last row is redundant, as $R_j$ is expressed in terms of the $z_j$ bosons. Here $S_{\alpha\alpha'}=\left(\begin{smallmatrix}1&0&0\\0&0&1\\0&1&0\end{smallmatrix}\right)$ and $\mathcal{K}$ denotes complex conjugation.}
    \label{tab: transformations Z2 QSL}
\end{table}
The transformation properties in Tab.~\ref{tab: transformations Z2 QSL} imply that the projective symmetry group (PSG) of the $\mathbb{Z}_2$ spin liquid defined by the actions~\eqref{eq: canted FM fermion action} and \eqref{eq: canted FM boson action} has the following nontrivial properties
\begin{equation}
    C_{4z}\sigma_d=-\sigma_d C_{4z}^{-1}\,,\hskip 2cm \sigma_d^2=-1\,.
\end{equation}
It is interesting to note that these are same transformation properties as those obeyed by a $d$-wave superconductor on the square lattice.

We also note that in this phase the chargon hopping $T^R_{jj'}$ takes the form
\begin{equation}\label{eq: canted FM chargon hoppings}
    T^R_{jj'}=\left(\begin{array}{cc} \langle z^*_j z_{j'} \rangle & \langle z^*_j\varepsilon z_{j'}^*\rangle \\ -\langle z_j\varepsilon z_{j'}\rangle & \langle z^*_{j'} z_{j} \rangle
    \end{array}\right)\propto \begin{cases}
        \left(
        \begin{smallmatrix}
            P & Q \\
            Q & P
        \end{smallmatrix}
        \right)\quad\text{if $j,j'=(d,p_x)$ or $(p_x,d)$}\\
        \left(
        \begin{smallmatrix}
            P & -Q \\
            -Q & P
        \end{smallmatrix}
        \right)\quad\text{if $j,j'=(d,p_y)$ or $(p_y,d)$}\\
    \end{cases}\,.
\end{equation}
Remembering that $Q/P=\tan(\theta/2)$ and comparing Eq.~\eqref{eq: canted FM chargon hoppings} with \eqref{eq: MF chargon action} and~\eqref{eq: canted FM fermion action}, we deduce \textit{a posteriori} that our approximation of replacing $t\to t_\mathrm{eff}$ is self-consistent. 
 
\subsection{Ferromagnetic Higgs phase}
In the ferromagnetic Higgs phase, we find $\avH{\mathfrak{j},\alpha}=H_0(0,0,1)$, with $\dmu_{p_x}=\dmu_{p_y}=\dmu_d\equiv\dmu$. The chargon action is obtained from action~\eqref{eq: canted FM fermion action} setting $\theta=0$, $H_d=H_{p_x}=H_{p_y}=H_0$ and $\dmu_{p_x}=\dmu_{p_y}=\dmu_d=\dmu$. Aside from the symmetries inherited from Eq.~\eqref{eq: canted FM fermion action}, the chargon action shows the additional symmetry $\psi_j\to e^{i\frac{\pi}{2}\sigma^3}\psi_j$, forcing all $\chi^{ss'}_{jj'}$ to be diagonal. This implies that the spinon action can be obtained from Eq.~\eqref{eq: canted FM boson action} by setting $Q=0$. We furthermore observe $\chi_\tau=\chi_0\,\mathrm{diag}(1,1,1,-1,-1,-1)$, with $\chi_0\simeq\frac{1}{2}$. This implies that in this case we have a U(1) spin liquid.

Within this gauge choice, the PSG acts trivially, that is, all symmetries can be implemented without resorting to a gauge transformation. 

It is worthwhile to remark that in this case the spinon action simply describes free bosons, which will realize a Bose-Einstein condensate at $T=0$, rendering the ferromagnetic Higgs phase always unstable to ferromagnetic long range order. 

\subsection{Ferrimagnetic Higgs phase}
Although we never find a ferrimagnetic Higgs phase at quarter filling within Hartree-Fock theory, we discuss it here as it emerges around half-filling $\nu=1/2$. In this phase we find $\avH{\mathfrak{j},\alpha}=H_\alpha (0,0,1)$ with $H_\alpha=\{H_d,-H_p,-H_p\}$, with $H_d, H_p>0$ and $\dmu_{p_x}=\dmu_{p_y}$. This phase can be obtained from the canted ferromagnetic Higgs phase by setting $\theta=\pi$. Therefore, the spinon action will be given by Eq.~\eqref{eq: canted FM boson action} with $P=0$. Note that with the help of the $\mathbb{Z}_2$ gauge transformation $z_d\to z_d$, $z_{p_x}\to z_{p_x}$, $z_{p_y}\to -z_{p_y}$, $\psi_d\to \psi_d$, $\psi_{p_x}\to \psi_{p_x}$, $\psi_{p_y}\to -\psi_{p_y}$, we can bring the $B_\bk^{\theta=\pi}$ (see Eq.~\eqref{eq: B matrix definition}) and $Q_\bk$ matrices (Eq.~\eqref{eq: Qk bosons SU(2)}) to the form
\begin{subequations}\label{eq: ferrimagnetic QSL gauge choice}
    \begin{align}
        B^{\theta=\pi}_\bk=&\left(
        \begin{array}{ccc}
            0 & 2t_\mathrm{eff}\cos\left(\frac{k_x}{2}\right) & 2t_\mathrm{eff}\cos\left(\frac{k_y}{2}\right)\\
            -2t_\mathrm{eff}\cos\left(\frac{k_x}{2}\right) & 0 & 0 \\
            -2t_\mathrm{eff}\cos\left(\frac{k_y}{2}\right)& 0 & 0 \\
        \end{array}
        \right)\,,\\
    Q_\bk=&\left(
    \begin{array}{ccc}
        0 & 2Q\cos\left(\frac{k_x}{2}\right) & 2Q\cos\left(\frac{k_y}{2}\right)\\
        -2Q\cos\left(\frac{k_x}{2}\right) & 0 & 0 \\
        -2Q\cos\left(\frac{k_y}{2}\right)& 0 & 0 \\
    \end{array}
    \right)\,.
    \end{align}
\end{subequations}
The action of the ferrimagnetic Higgs phase is invariant under the transformations $\psi_{\mathfrak{j},\alpha}\to e^{i\phi_\alpha \frac{\sigma^3}{2}}\psi_{\mathfrak{j},\alpha}$,  $z_{\mathfrak{j},\alpha}\to e^{-i\frac{\phi_\alpha}{2}}z_{\mathfrak{j},\alpha}$, where $\phi_\alpha=\{\phi,-\phi,-\phi\}$ and $\phi$ is an arbitrary angle. This observation implies that the invariant gauge group (IGG) of this state is U(1), rendering the ferrimagnetic Higgs phase a U(1) spin liquid. Within the gauge choice in Eq.~\eqref{eq: ferrimagnetic QSL gauge choice}, the PSG of this phase is trivial.

\subsection{Spin gap in the canted ferromagnetic Higgs phase}
\begin{figure}[t]
    \centering
    \includegraphics[width=0.5\linewidth]{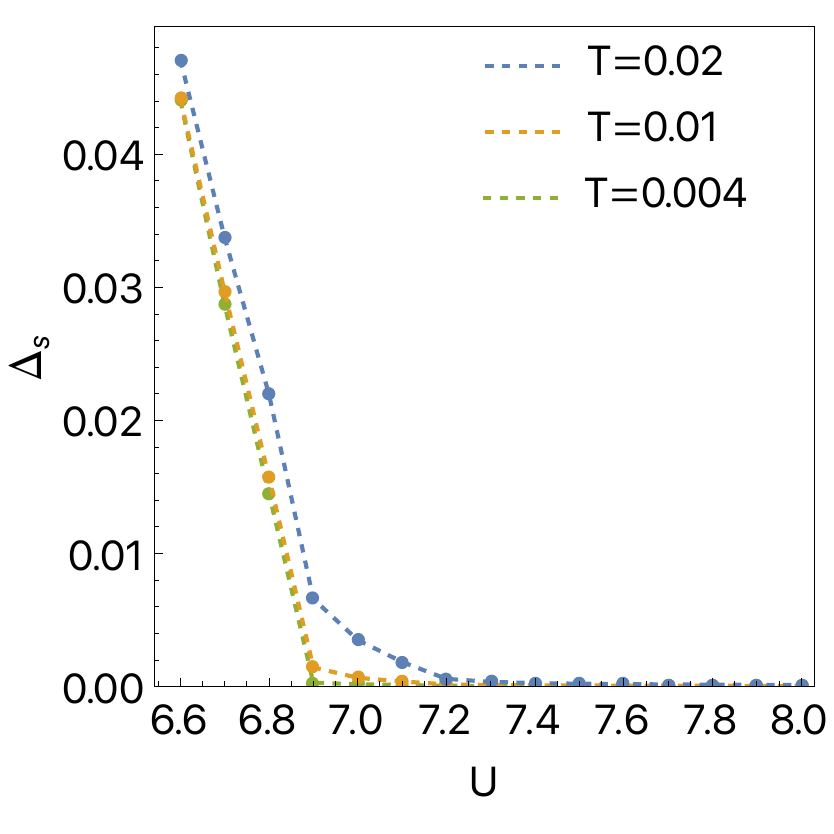}
    \caption{Spin gap computed within the SU(2) gauge theory by forcing a canted ferromagnetic Higgs phase for three different values of the temperature (in units of the renormalized hopping $t_\mathrm{eff}$). We obtain a spin gapped phase only for $U\lesssim 6.9 \,t_\mathrm{eff}$.}
    \label{fig: SU2 spin gap}
\end{figure}
In this section, we compute the spin gap at quarter filling in the canted ferromagnetic phase. This can be done by determining the mean-field values of the Lagrange multipliers $\bar{\lambda}_d$ and $\bar{\lambda}_p$ in Eq.~\eqref{eq: Pk bosonic action canted FM Higgs} by imposing that the constraint on the bosonic spinons is fulfilled on average:
\begin{equation}
    \langle z^*_j z_j\rangle =1\,,
\end{equation}
where the average is computed using action~\eqref{eq: canted FM boson action}. The remaining parameters, that is, $P$, $Q$, $\chi_{d,\tau}$ and $\chi_{p,\tau}$ are computed within Hartree-Fock theory for the chargons, using the renormalized chargon hopping $t_\mathrm{eff}$ as a free parameter. Once all parameters in the bosonic action~\eqref{eq: canted FM boson action} are known, one can compute the spectrum around $\bk=0$, where a minimum in the lowest band occurs, and the value of the lowest band corresponds to the spin gap. 

In Fig.~\ref{fig: SU2 spin gap} we show a calculation of the spin gap $\Delta$ within the SU(2) gauge theory at quarter filling as a function of the Hubbard $U$ at three different temperatures, always forcing a canted ferromagnetic Higgs phase. All quantities are given in units of the effective hopping $t_\mathrm{eff}<t$. We observe that for $U\gtrsim6.9\, t_\mathrm{eff}$ the spin gap takes very small values, eventually vanishing at $T=0$, signaling the appearance of long range canted ferromagnetic order in the ground state. Differently, for $U\lesssim 6.9\, t_\mathrm{eff}$, the spin gap remains sizable down to the lowest temperatures, signaling the formation of a spin gapped ground state, corresponding to a $\mathbb{Z}_2$ spin liquid with gapless fermionic chargons forming an almost flat band at the Fermi level (see Sec.~\ref{subsec: spectrum canted FM}). 

\subsection{$(q,0)$ Spiral Higgs phase}
The spiral Higgs phase is characterized by a Higgs condensate of the form
\begin{equation}
    \avH{\mathfrak{j},\alpha} = 
    \begin{cases}
        H_d\, (\sin(\bQ\cdot \bR_{\mathfrak{j},d}),0,\cos(\bQ\cdot \bR_{\mathfrak{j},d}))&\alpha=d\,, \\
        -H_{p_x}\, (\sin(\bQ\cdot \bR_{\mathfrak{j},p_x}),0,\cos(\bQ\cdot \bR_{\mathfrak{j},p_x}))&\alpha=p_x\,, \\
        H_{p_x}\, (\sin(\bQ\cdot \bR_{\mathfrak{j},p_y}),0,\cos(\bQ\cdot \bR_{\mathfrak{j},p_y}))&\alpha=p_y\,, 
    \end{cases}
\end{equation}
where $\bQ=(q,0)$ and $H_\alpha>0$ for each $\alpha$. As done in the previous sections, we perform an SU(2) gauge transformation such that the Higgs condensate in the new basis takes the form $\avH{j}\propto (0,0,1)$. This is achieved using the SU(2) matrix $V_j=e^{-i\frac{\phi_j}{2}\sigma^2}$, with $\phi_{\mathfrak{j},d}=\bQ\cdot \bR_{\mathfrak{j},d}$, $\phi_{\mathfrak{j},p_x}=\bQ\cdot \bR_{\mathfrak{j},p_x}+\pi$ and $\phi_{\mathfrak{j},p_y}=\bQ\cdot \bR_{\mathfrak{j},p_y}$. In this gauge the chargon action reads
\begin{subequations}\label{eq: spiral Higgs fermion action}
    \begin{align}
        &\mathcal{S}_\psi[\psi,\bar{\psi}]=-T\sum_{\omega_n}\int_\bk\,\Psi^\dagger_{k} \left[i\omega_n\mathbbm{1}_6-\left(
        \begin{array}{c|c}
            A^\bQ_\bk - H & B_\bk^\bQ \\ \hline
            (B_\bk^\bQ)^\dagger & A^\bQ_\bk+H
        \end{array}
        \right)\right]\Psi_{k}\,,\\
        &A^\bQ_\bk=\left(
        \begin{array}{ccc}
            -(\mu+\delta\mu_d) & 2i\,t_\mathrm{eff}\cos\left(\frac{q}{4}\right)\cos\left(\frac{k_x}{2}\right) & -2\,t_\mathrm{eff}\cos\left(\frac{k_y}{2}\right)\\
            -2i\,t_\mathrm{eff}\cos\left(\frac{q}{4}\right)\cos\left(\frac{k_x}{2}\right) & -(\mu+\delta\mu_{p_x}) & 0 \\
            -2\,t_\mathrm{eff}\cos\left(\frac{k_y}{2}\right) & 0 & -(\mu+\delta\mu_{p_y})
        \end{array}
        \right)\,,\\
        &B^\bQ_\bk=\left(
        \begin{array}{ccc}
            0 & 2\,t_\mathrm{eff}\sin\left(\frac{q}{4}\right)\sin\left(\frac{k_x}{2}\right) & 0\\
            2\,t_\mathrm{eff}\sin\left(\frac{q}{4}\right)\sin\left(\frac{k_x}{2}\right) & 0 & 0 \\
            0 & 0 & 0
        \end{array}
        \right)\,,\\
        &H=\mathrm{diag}(H_d,H_{p_x},H_{p_y})\,. 
    \end{align}
\end{subequations}
Evaluating the coefficients in Eq.~\eqref{eq: MF spinon action} from action~\eqref{eq: spiral Higgs fermion action}, we obtain the following action for the bosonic spinons
\begin{subequations}\label{eq: spiral spinon action}
    \begin{align}
    \mathcal{S}_z[z,z^*]=&-T\sum_{\Omega_n}\int_\bk\,Z^\dagger_{k} \left[i\chi_\tau\Omega_n-\left(
        \begin{array}{c|c}
            P_\bk & Q_\bk \\ \hline
            Q_\bk^\dagger &P_{-\bk}^T
        \end{array}
        \right)\right]Z_{k}\,,\\
        P_\bk=&\left(
        \begin{array}{ccc}
            -\bar{\lambda}_d & 2iP_x\sin\left(\frac{k_x}{2}\right) & 2P_y\cos\left(\frac{k_y}{2}\right)\\
            -2iP_x\sin\left(\frac{k_x}{2}\right) & -\bar{\lambda}_{p_x} & 0 \\
            2P_y\cos\left(\frac{k_y}{2}\right)& 0 & -\bar{\lambda}_{p_y} \\
        \end{array}
        \right)\,,\\
        Q_\bk=&\left(
        \begin{array}{ccc}
            0 & +2Q\cos\left(\frac{k_x}{2}\right) & 0\\
            -2Q\cos\left(\frac{k_x}{2}\right) & 0 & 0 \\
            0 & 0 & 0 \\
        \end{array}
        \right)\,,\label{eq: Qk bosons SU(2) spiral}\\
        \chi_\tau=&\mathrm{diag}(\chi_{d,\tau},\chi_{p_x,\tau},\chi_{p_y,\tau},-\chi_{d,\tau},-\chi_{p_x,\tau},-\chi_{p_y,\tau})\,,
    \end{align}
\end{subequations}
where, similarly to the previous sections, $\bar{\lambda}_d$, $\bar{\lambda}_{p_x}$ and $\bar{\lambda}_{p_y}$ are chosen to enforce $\langle z^*_j z_j\rangle=1$.
We observe that if the coefficients of the spinon action above are calculated from a converged Hartree-Fock solution, then $\frac{P_x}{P_y}=\cos\left(\frac{q}{4}\right)$ and $\frac{Q}{P_x}=\tan\left(\frac{q}{4}\right)$. This fact ensures that when computing the chargon hopping matrix $T^R_{jj'}$ from Eq.~\eqref{eq: spiral spinon action}, this will have the same structure as in Eq.~\eqref{eq: spiral Higgs fermion action}, making the initial assumption of replacing $t\to t_\mathrm{eff}$ self-consistent. 

Analyzing Eq.~\eqref{eq: spiral spinon action}, we can see that because the spinon "pairing terms" are nonvanishing only on $x$-bonds, this theory describes a $\mathbb{Z}_2$ spin liquid that spontaneously breaks the lattice $C_{4v}$ point symmetry group. Furthermore, we observe that the spinon dispersion exhibits two degenerate minima at $\bk=(\pm k_0,0)$ in the lowest band, where $k_0$ is approximately $q/2$ if $P_x$, $P_y$, and $Q$ are computed from a converged Hatree-Fock solution and if $\bar{\lambda}_\alpha$ are such that the spinons are gapless (implying long range spiral order) or nearly gapless.

\section{Schwinger boson theory}
\label{sec:sb}

We start this section by introducing a $t$-$J$ Hamiltonian, corresponding to the large-$U$ limit of Eq.~\eqref{eq: Hubbard model}
\begin{equation}
    H=-t P\sum_{\langle j,j'\rangle,\sigma} c^\dagger_{j,\sigma}c_{j',\sigma}P + J\sum_{\langle j,j'\rangle} \vec{S}_j \cdot \vec{S}_{j'},
\end{equation}
where $t$ is the hopping term and $J>0$ is a Heisenberg interaction between the spins, and $P$ are the projection operators on the basis with no double occupancy. The fermions described by $c_{j,\sigma}$ are subject to the constraint $\sum_\sigma c^\dagger_{j,\sigma}c_{j,\sigma}\leq 1$. To enforce this constraint, we use the  decomposition (\ref{cbf}) of the fermionic operator $c_{i \sigma}=b_{i \sigma}f^{\dag}_i$ in terms of the Schwinger boson and the fermionic spinless chargon, with a constraint at each site $b_{i \sigma}^\dag b_{i \sigma}+f^{\dag}_i f_i=1$. By introducing auxiliary fields and treating them with the saddle-point approximation, we arrive at the following mean-field Hamiltonian:

\begin{equation}
    H_{mf}= \sum_{\langle j,j'\rangle,\sigma, \sigma'} \left(-Q_{j,j'} \epsilon_{\sigma \sigma'} b_{j \sigma}^\dag b_{j' \sigma'}^\dag  +
  (  P_{j,j'}+ R_{j,j'}) \delta_{\sigma \sigma'}  b_{j \sigma}^\dag b_{j' \sigma'}+\frac{2 t}{J}P_{j,j'}f^{\dag}_j f_{j'}+ c. c.      
    \right)+\sum_j \lambda_j( b_{j \sigma}^\dag b_{j \sigma}+f^{\dag}_j f_j-1).
\end{equation}
The mean-field parameters are defined in the following way:
\begin{equation}
\begin{dcases}
  &  Q_{j,j'}=\frac{J}{2}\epsilon_{\sigma \sigma'} \langle b_{j \sigma} b_{j' \sigma'} \rangle\\
   &  P_{j,j'}=\frac{J}{2}\delta_{\sigma \sigma'} \langle b_{j \sigma}^\dag b_{j' \sigma'} \rangle\\
  & R_{j,j'}=t  \langle f_j^\dag f_j' \rangle .\\ 
\end{dcases}
\end{equation}

\begin{figure}[t!]
    \centering
    \includegraphics[width=0.4\linewidth]{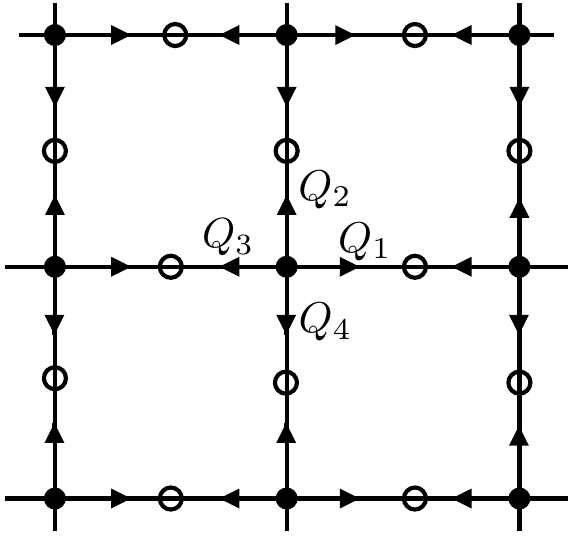}
    \caption{The directions of the $Q_{j,j'}$ field inside the unit cell of the Lieb lattice. }
    \label{fig:direction}
\end{figure}

In the most general case, the mean-field parameters depend on the position indexes $(j,j')$. However, in the rest of the paper, we assume the system does not break any translational invariance. This allows us to drop indexes. We also note that $Q_{j,j'}=-Q_{j',j}$ so we choose the directions of the field within the unit cell as shown in Fig. \ref{fig:direction}. There are three distinct Lagrange multipliers $\lambda_d, \lambda_{p_x}, \lambda_{p_y}$ on each site in the unit cell. 

To proceed further we go to a Fourier space and introduce the basis:
\begin{equation}
\begin{dcases}
& B_k=\left(b_{d \uparrow}(k), b_{p_{x} \uparrow}(k), b_{p_{y} \uparrow}(k), b^{\dagger}_{d \downarrow}(-k),b^{ \dagger}_{p_x \downarrow}(-k),b^{ \dagger}_{p_y \downarrow}(-k) \right)  \\
 &\psi_k= \left( f_d(k), f_{p_x}(k), f_{p_y}(k)\right)\\
\end{dcases}
\end{equation}
In the chosen basis the boson Hamiltonian is given by $6\times 6$ matrix
 \begin{equation}
 H_k^b= \bar{B}_k\left(
     \begin{array}{cc}
          \lambda +P_k+R_k& Q_k \\
          Q^{\dagger}_k& \lambda +P^*_{-k}+R^*_{-k}
     \end{array}
     \right)B_k,
 \end{equation}
with 

\begin{subequations}
    \begin{align}
         Q_k= &\left(
         \begin{array}{ccc}
             0&- Q_1 e^{i k_x/2}-Q_3 e^{-i k_x/2} &- Q_2 e^{i k_y/2}-Q_4 e^{-i k_y/2} \\
              Q_1 e^{-i k_x/2}+Q_3 e^{i k_x/2}& 0&0\\
              Q_2 e^{-i k_y/2}+Q_4 e^{i k_y/2}&0&0\\
         \end{array}
         \right),\\
     P_k= &\left(
         \begin{array}{ccc}
             0& P^*_1 e^{i k_x/2}+P^*_3 e^{-i k_x/2} & P_2^* e^{i k_y/2}+P_4^* e^{-i k_y/2} \\
              P_1 e^{-i k_x/2}+P_3 e^{i k_x/2}& 0&0\\
              P_2 e^{-i k_y/2}+P_4 e^{i k_y/2}&0&0\\
         \end{array}
         \right),
     \end{align}
 \end{subequations}
and $\lambda= diag(\lambda_d, \lambda_{p_x}, \lambda_{p_y})$. $R_k$ has the same form as the $P_k$ matrix with $P \rightarrow R$. The fermion Hamiltonian is a $3 \times 3 $ matrix 
 \begin{equation}
 H^f_k=\frac{2t}{J } \bar{\psi}_k\left(
     \begin{array}{ccc}
         \lambda_d+\mu& P^*_1 e^{i k_x/2}+P^*_3 e^{-i k_x/2} & P_2^* e^{i k_y/2}+P_4^* e^{-i k_y/2} \\
          P_1 e^{-i k_x/2}+P_3 e^{i k_x/2}& \lambda_{p_x}+\mu&0\\
          P_2 e^{-i k_y/2}+P_4 e^{i k_y/2}&0&\lambda_{p_y}+\mu \\
     \end{array}
     \right) \psi_k.
 \end{equation}

We also assume that there is no nematic phase, so mean-field parameters in different directions have the same absolute value: $|Q_1|=|Q_2|=|Q_3|=|Q_4|$ and the same for $P$ and $R$. This assumption further implies that $\lambda_{p_x}=\lambda_{p_y}=\lambda_p$. All $P$ and $R$ values are taken to be positive without loss of generality (they can be made positive by a gauge transformation). With $Q_1=Q_3=Q$ there are two distinct combinations $Q_2=Q_4=Q$ and $Q_2=Q_4=-Q$. We find that the second case corresponds to the canted order described earlier and has a lower energy, so below we only consider the latter case.

The self-consistency equations are:

\begin{equation}
\begin{dcases}
   & Q=\sum_k \left( \langle b_{d, \uparrow}(k) b_{p_x, \downarrow}(-k) \rangle e^{-i k_x/2}-\langle b_{d, \downarrow}(-k) b_{p_x, \uparrow}(k) \rangle e^{i k_x/2} \right) \\
   &    P= \sum_k \left( \langle b^\dag_{d \uparrow}(k) b_{p_x \uparrow}(k) \rangle e^{i k_x/2}+ \langle b_{d \downarrow}^\dag(-k) b_{p_x \downarrow}(-k) \rangle e^{-i k_x/2} \right)\\
   & R=\sum_k \langle f^{\dag}_d(k) f_{p_x}(k)\rangle e^{i k_x/2}.\\
    \end{dcases}
    \label{eqn:self_const1}
\end{equation}
To find Lagrange multipliers $\lambda_d$,$\lambda_p$ we impose the constraint from the fractionalization and also tune the chemical potential $\mu$ to ensure the system is at quarter filling: 
 \begin{equation}
 \begin{dcases}
&n_d^f+n_d^b=\sum_k \langle f^{\dag}_{d}(k) f_{d}(k) \rangle+\sum_{k,\alpha=\up, \down} \langle b^{\dag}_{d, \alpha}(k) b_{d, \alpha}(k) \rangle=1\\ 
&n_{p_x}^f+n_{p_x}^b=\sum_k \langle f^{\dag}_{p_x}(k) f_{p_x}(k) \rangle+\sum_{k,\alpha={\up, \down}} \langle b^{\dag}_{p_x, \alpha}(k) b_{p_x, \alpha}(k) \rangle=1\\ 
 &    n^f=\sum_{a=(d,p_x,p_y)}\sum_k \langle f^{\dag}_{a}(k) f_{a}(k) \rangle=\frac{3}{2}. \\
     \end{dcases}
     \label{eqn:self_const2}
 \end{equation}

\begin{figure}
    \centering
    \includegraphics[width=0.45\linewidth]{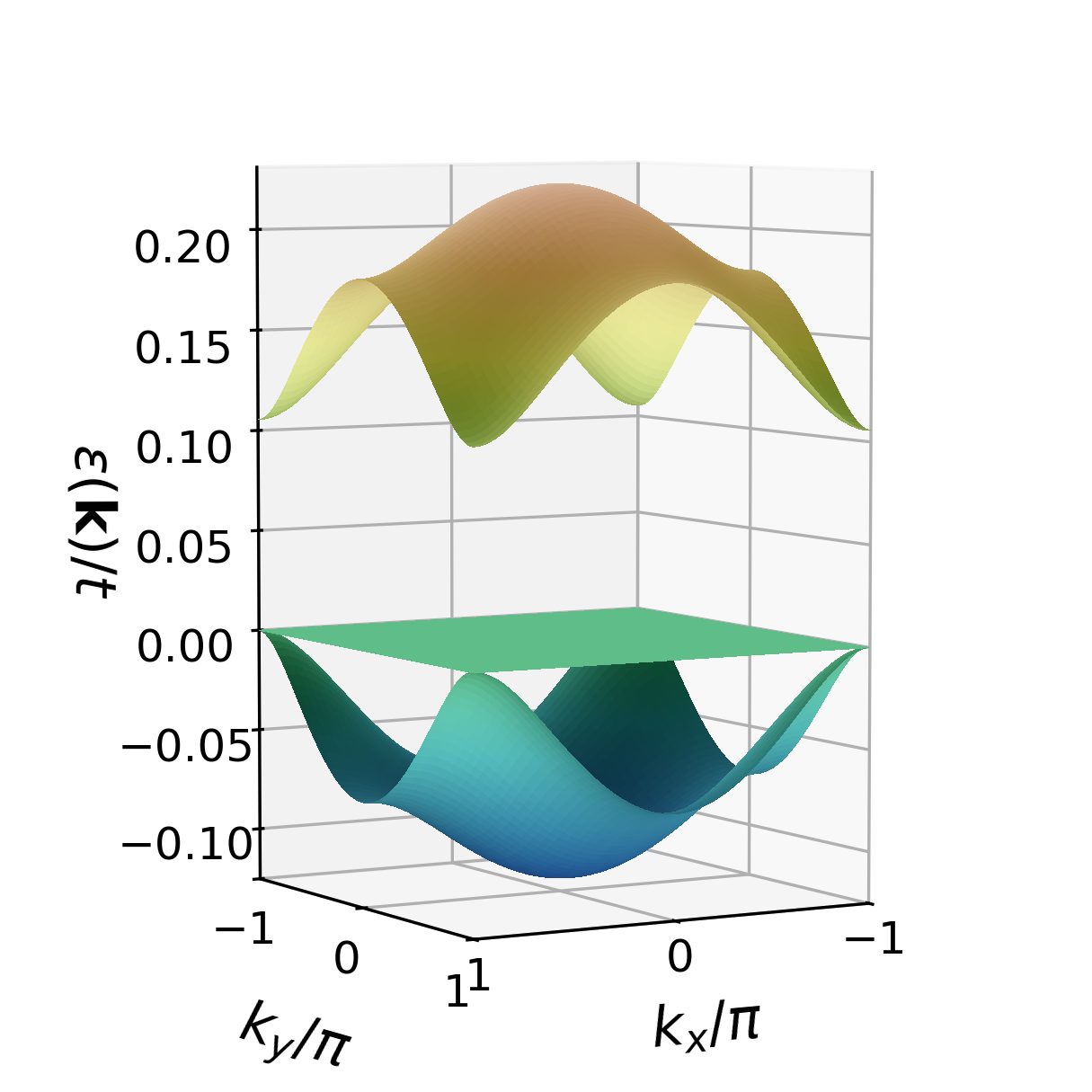}
    \caption{Fermionic bands at quarter filling as computed within Schwinger boson theory. The parameters are $L_x=L_y=20$, $T=0.02$, $J=2$.}
    \label{fig: holon bands}
\end{figure}

The fermionic Hamiltonian can be easily diagonalized via the transformation $f_a(k)=U(k)^{ab} \tilde{f}_b(k)$, where $U$ is the unitary matrix made from the eigenvectors of $H^f(k)$.
The fermionic spectrum, shown in Fig.~\ref{fig: holon bands}, looks identical to the non-interacting case of spinless electrons with the renormalized coefficients and different on-site potential for $d$ and $p$ sites. Comparing Fig.~\ref{fig: holon bands} to Fig.~\ref{fig: Lieb lattice}(b), we observe that the different chemical potentials on $d$ and $p$ sites gap out one of the two dispersing bands from the flat band, which is now quadratically touched only by the other dispersing band. Since the fermions are spinless, the Fermi energy should be close to flat bands at quarter filling instead of half filling. This observation could naturally explain the cold-atom observations \cite{Lebrat24}.

The bosonic Hamiltonian is more subtle to diagonalize. Using the similar transformation $B_a(k)=T(k)^{ab} \tilde{B}_b(k)$ and imposing the same commutation relation to hold for $\tilde{B}$, one concludes that $T$ is a pseudounitary matrix: $T \sigma_3 T^{\dag}=\sigma_3$, where $\sigma_3=diag(1,1,1,-1,-1,-1)$. Therefore, the diagonal Hamiltonian $H'$ is related to the original Hamiltonian by the following similarity transformation $\sigma_3 H'=T^{-1} (\sigma_3 H) T$. To find the eigenvalues of $H'$ one could diagonalize the bosonic Hamiltonian $\sigma_3 H$, since similarity transformations preserve the eigenvalues. We note that the eigenvectors of $\sigma_3 H$ are the columns of the transformation matrix $T$. Hence, we are able to construct the matrix $T$ by taking the corresponding eigenvectors of $\sigma_3 H$, and normalizing them to satisfy the pseudo-unitarity condition.

In the end, we have 6 unknown parameters $Q,P,R,\lambda_d,\lambda_p,\mu$ and 6 self-consistency equations \ref{eqn:self_const1}, \ref{eqn:self_const2} which allows us to solve them for different couplings $J$ and temperatures $T$ (in the rest of the paper we set $t=1$ and measure $J$ and $T$ in the units of $t$). We chose $L_x=L_y=20$ to perform the summation over the Brillouin zone and solve the self-consistency equations using Broyden's method. 

 \begin{figure}[H]
  \begin{minipage}[h]{0.45\linewidth}
  \center{\includegraphics[width=1\linewidth]{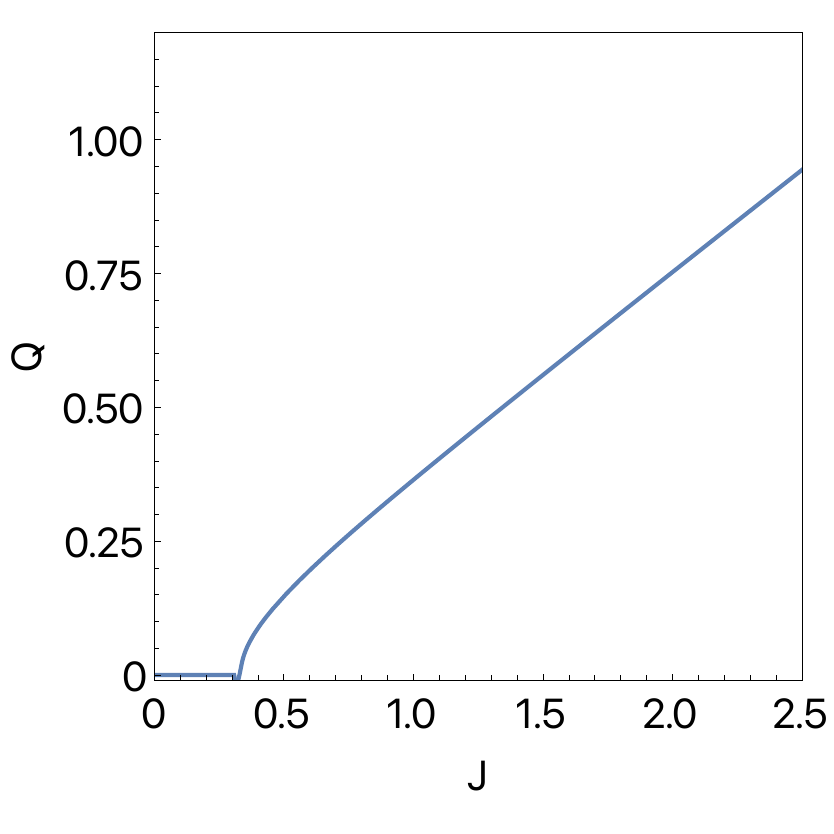}}
  \end{minipage} 
       \hfill 
\begin{minipage}[h]{0.45\linewidth}
  \center{\includegraphics[width=1\linewidth]{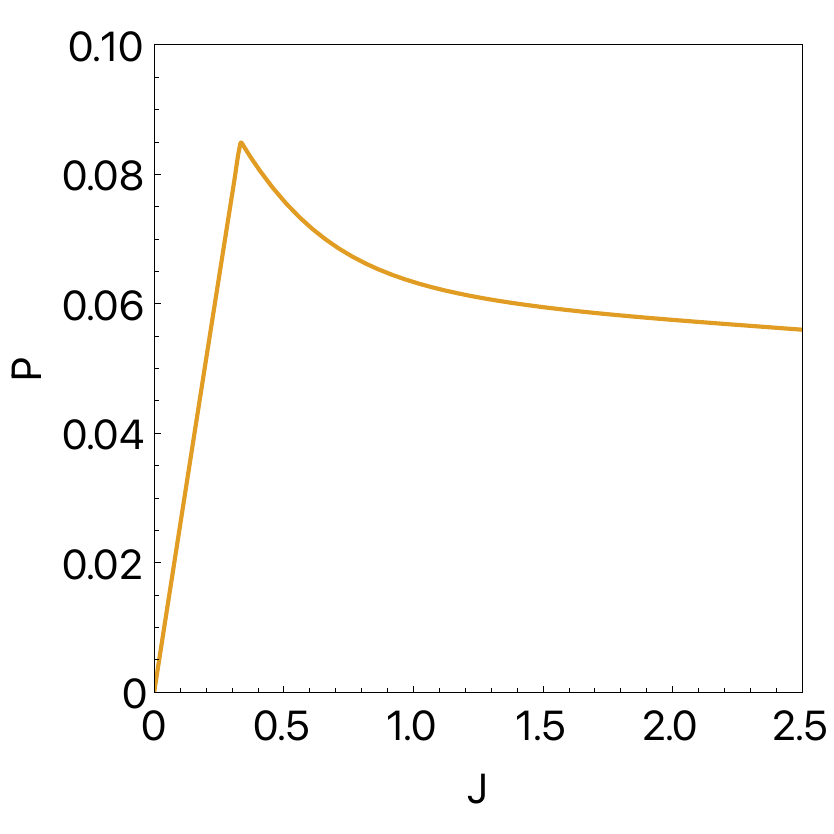}}
  \end{minipage} 
\caption{The mean-field parameters as a function of $J$. $L_x=L_y=20$ and $T=0.02$.}
\label{fig:mean_fields}
\end{figure}

Fig. (\ref{fig:mean_fields}) shows that for small $J<J_{c1} \approx 0.34$ we find a phase with $Q=0,P \neq 0$, while for larger $J>J_{c1}$ there is a phase with both $Q\neq 0,P \neq 0$. We can understand the nature of the phases better by drawing a classical analogy. Let us consider the classical spin vectors in the configuration depicted in Fig. (\ref{fig:canted}) and express them in terms of classical fields:
\begin{equation}
   \vec{S}_d = d^* \vec{\sigma} d= m_d \hat{z}, \quad \vec{S}_{p_x}=p_x^* \vec{\sigma} p_x= m_p( \cos \theta \hat{z}+\sin \theta \hat{x}), \quad  \vec{S}_{p_y}=p_y^* \vec{\sigma} p_y= m_p( \cos \theta \hat{z}-\sin \theta \hat{x})
\end{equation}
After solving for the classical fields, we can express mean-field parameters in terms of canting angle and magnetizations:
\begin{equation}
\begin{split}
 &   P_x=d^* p_x=\sqrt{m_d m_p} \cos \theta/2 , \quad P_y=d^* p_y=\sqrt{m_d m_p} \cos \theta/2\\
 &Q_x=d \epsilon p_x = \sqrt{m_d m_p} \sin \theta/2 , \quad Q_y=d \epsilon p_y = -\sqrt{m_d m_p} \sin \theta/2 \\
\end{split}
\end{equation}

\begin{figure}[H]
  \begin{minipage}[h]{0.45\linewidth}
  \center{\includegraphics[width=1\linewidth]{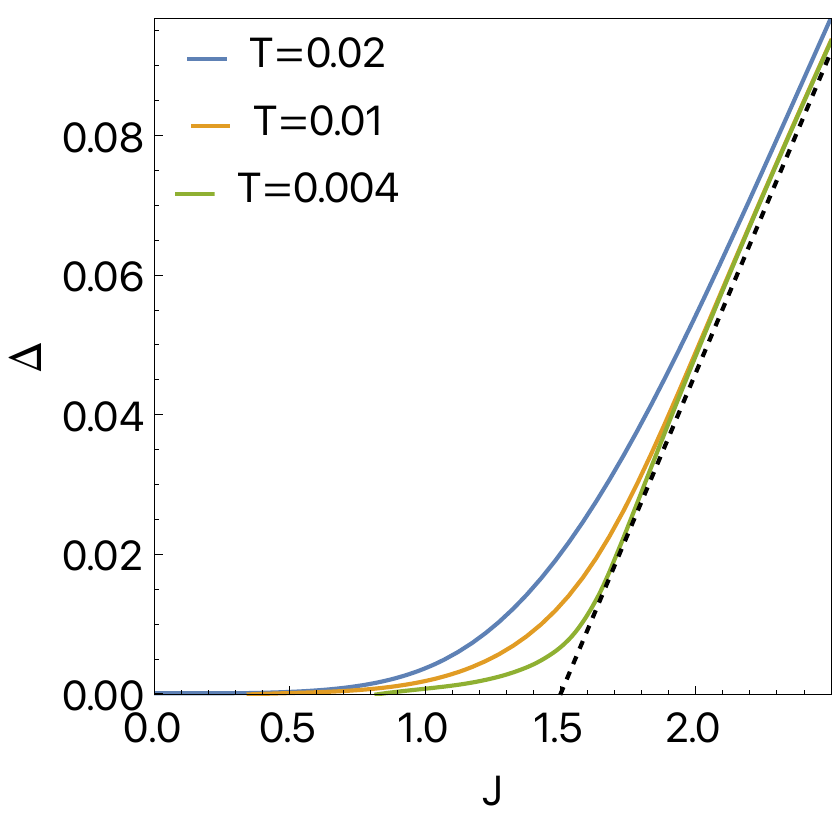}}
  \\a)
  \end{minipage} 
       \hfill 
\begin{minipage}[h]{0.45\linewidth}
  \center{\includegraphics[width=1\linewidth]{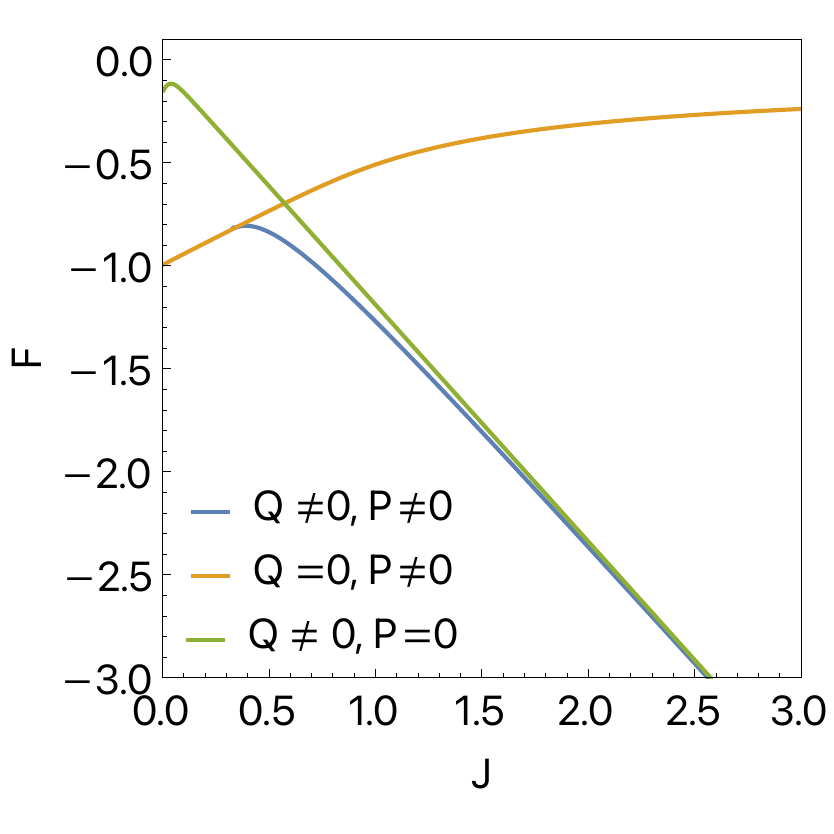}}
  \\b)
  
  \end{minipage} 
\caption{a) The bosonic spin gap as a function of $J$ for different temperatures. Black dashed line correspond to interpolation to zero temperature. b) The free energy of three different phases.}
\label{fig:gap_energy}
\end{figure}

Using this analogy, we conclude that the phase with $Q=0,P \neq 0$ corresponds to the ferromagnetic phase with $\theta=0$ when the Schwinger bosons condense. The similar phase is predicted by the Hartree-Fock theory.
The phase with $Q\neq 0,P \neq 0$ corresponds to a canted phase, with canting angle given by the equation: $\tan \theta/2 = P/Q$. To further understand the nature of the canted phase, we computed a spin gap (which is given by the bosonic spectrum, since bosons carry charge) and checked how it scales with temperature. Fig. \ref{fig:gap_energy}(a) demonstrates the existence of two regions $J<J_{c2}\approx 1.5$ where the gap goes to zero as we lower the temperature and  $J>J_{c2}$ where the gap remains finite. The first region corresponds to the previously mentioned canted phase, the second region is $\mathbb{Z}_2$ spin liquid.

Apart from the three phases above, we also observed the phase with $Q\neq 0, P=0 $ and finite spin gap. To compare which of the phases is energetically favorable, we computed the free energies of all phases. The expression for the free energy looks as follows:

\begin{equation}
    F=E_{mf}+\frac{8 Q^2}{J}-\frac{8 P^2}{J}-\frac{16 P R}{J}-2\lambda_d-4 \lambda_p-3\mu n_e.
\end{equation}
The expression can be derived by properly introducing mean-field parameters using the Hubbard–Stratonovich transformation and taking their value at the saddle point. Bosonic and fermionic energies are given by the usual formulas:
\begin{equation}
   E_{mf}=\sum_{a=1,2,3} 2 \sum_k \left(E_a(k)+ 2T \log \left(1-e^{-E_a(k)/T}\right) -T \log \left(1+e^{-\epsilon_a(k)/T}\right)\right),
\end{equation}
where $E_a(k)$ is bosonic dispersion relation and $\epsilon_a(k)$ is the fermionic dispersion relation. We checked that the mean-field parameters obtained earlier correspond to extremum values of the free energy $F$.
Fig. \ref{fig:gap_energy} (b) shows the free energy of all discussed phases. We see that the phase with $P=0$ is always higher in energy so it is never realized. The ferromagnetic phase at small $J$ turns into a canted phase at $J>J_{c1}$, since it becomes more energetically favorable. 

\section{DMRG}
\label{sec:dmrg}
In this section, we analyze the Hubbard model and the $t$-$J$ model on the Lieb lattice using DMRG (density matrix renormalization group) method. The complexity of the method scales exponentially with $L_y$, so we study the model on a cylinders with $L_y=2$ unit cells in the $y$ direction. Since $L_y$ is small, the system is effectively quasi one-dimensional and we can apply the Luttinger liquid theory to describe the effects of interaction. For example, for commensurate fillings $n=1/2,1/3,1/6$ the Umklapp processes are relevant and the system acquires a charge gap. The detailed analysis of the phase diagram will be done in the forthcoming paper, while here we include several crucial observations.

 \begin{figure}[H]
  \begin{minipage}[h]{0.45\linewidth}
  \center{\includegraphics[width=1\linewidth]{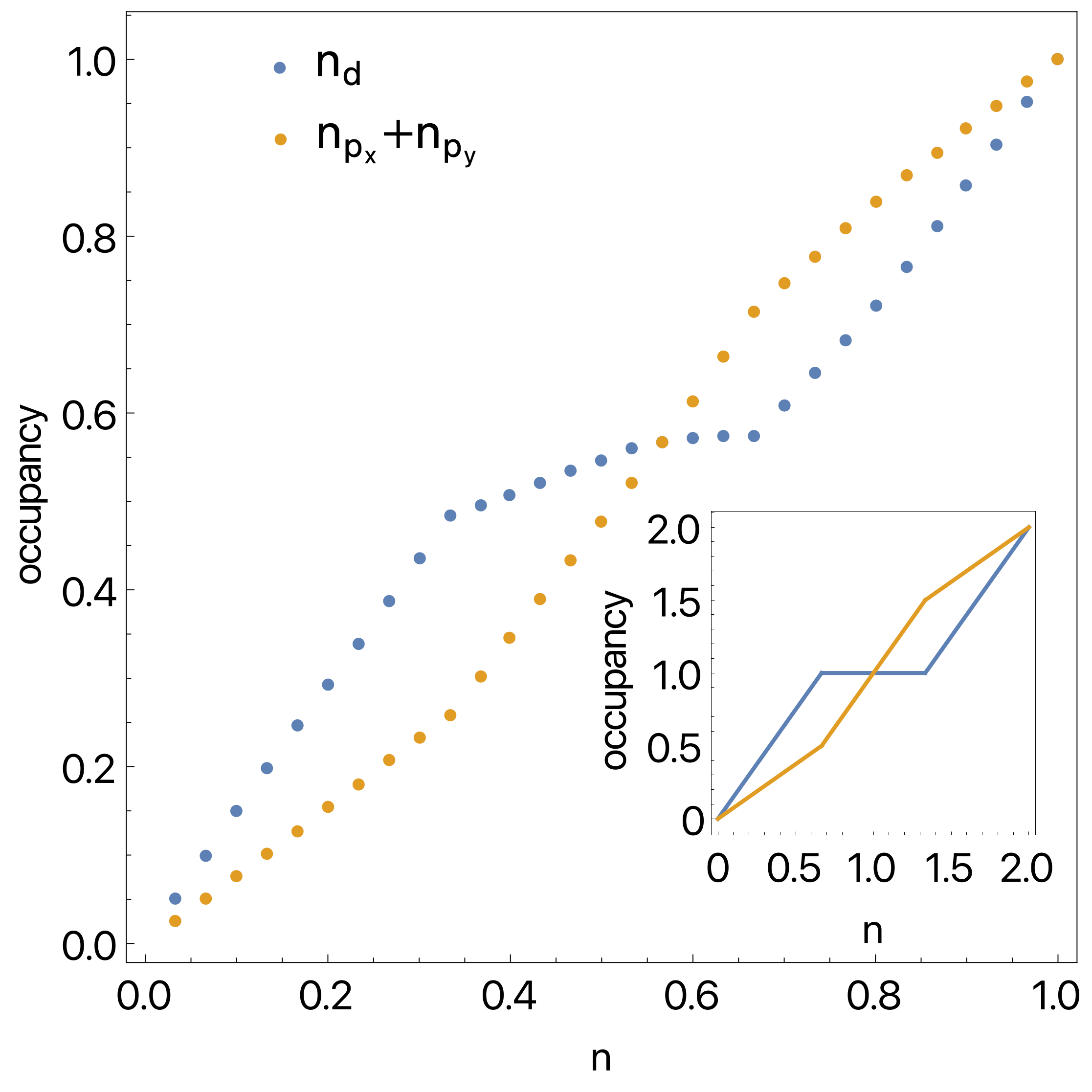}}
  \\a) $U=16$
  \end{minipage} 
       \hfill 
\begin{minipage}[h]{0.45\linewidth}
  \center{\includegraphics[width=1\linewidth]{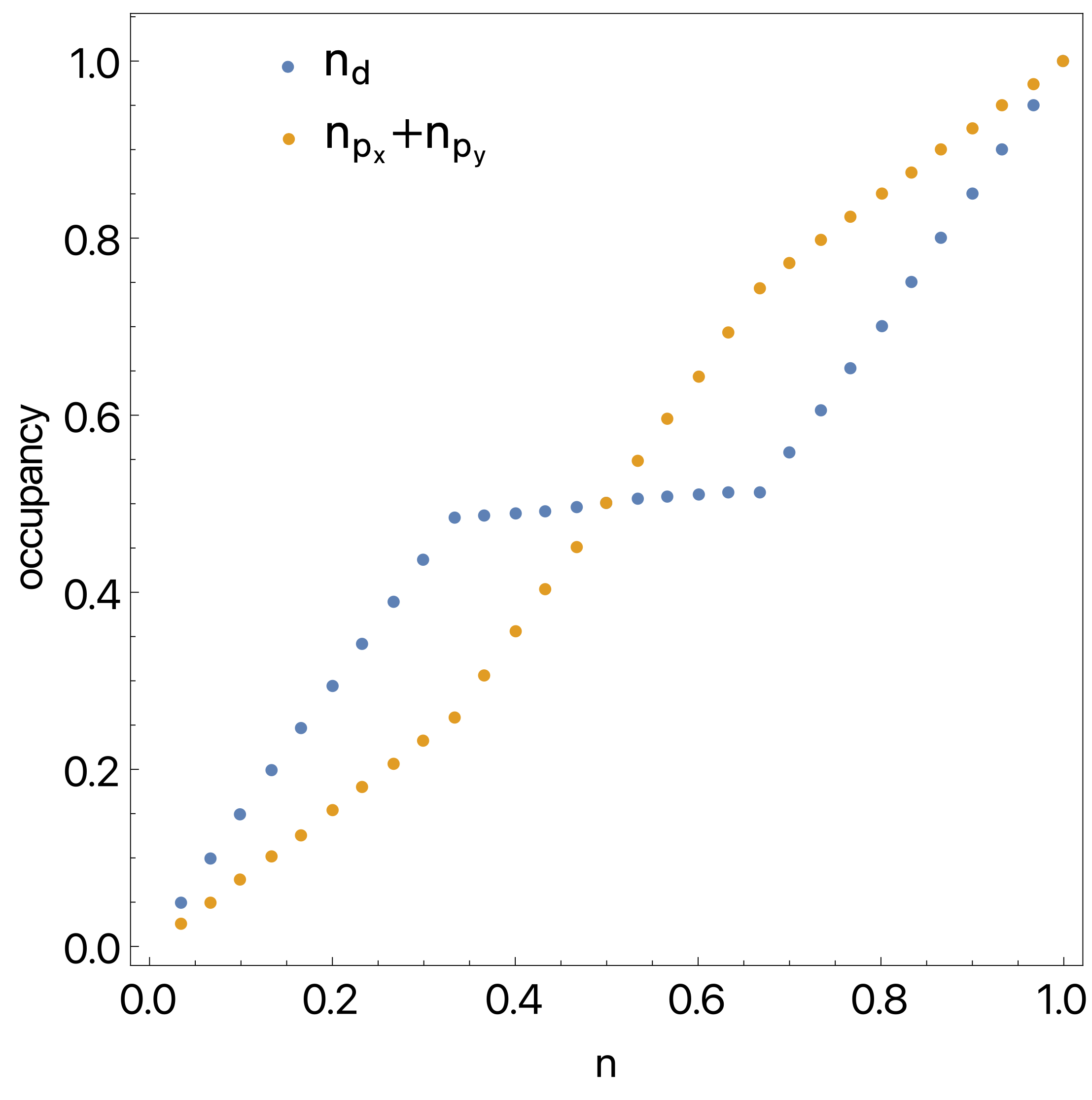}}
  \\b) $U=\infty$
  \end{minipage}

\caption{The total density on $n_d$ sites and on $n_{p_x}+n_{p_y}$ sites as the function of filling for the Hubbard model with $U=16$. Inset shows non-interacting case. Parameters: $L_x \times L_y= 10 \times2$, $U=16$, $t=1$, $D=2000$. }
\label{fig:DMRG}
\end{figure}

Fig. \ref{fig:DMRG}a shows the occupation of $n_d$ sites and of $n_{p_x}+n_{p_y}$ separately as a function of total filling. One can compare it to the non-interacting case, where the plot is almost scaled twice. The comparison is even more striking, see Fig. \ref{fig:DMRG}b,  for the $t$-$J$ model with $J=0$, which is equivalent to Hubbard model with infinite interactions. A similar behaviour was observed in the experimental paper~\cite{Lebrat24}. One of the possible scenarios to explain the doubling was presented above: it includes development of the magnetic order and partial spin polarization. For example: in the fully spin-polarized system one half of the electrons is completely gapped out, and there is effectively one electron per site.
 \begin{figure}[H]
  \begin{minipage}[h]{0.45\linewidth}
  \center{\includegraphics[width=1\linewidth]{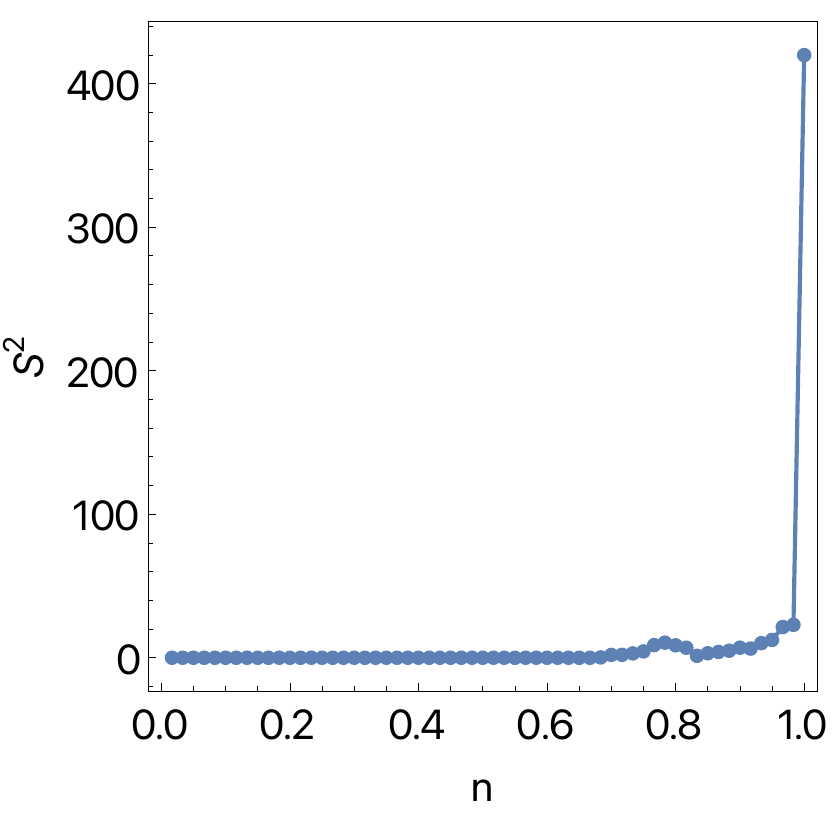}}
  \\a) $U=16$
  \end{minipage} 
       \hfill 
\begin{minipage}[h]{0.45\linewidth}
  \center{\includegraphics[width=1\linewidth]{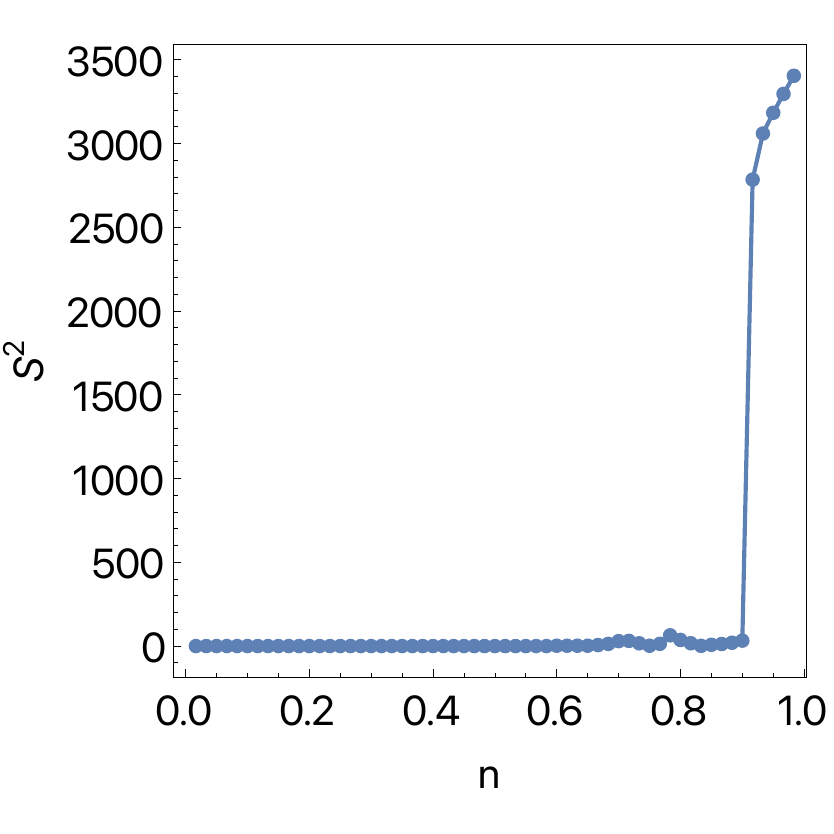}}
  \\b) $U=\infty$
  \end{minipage} 
  
\caption{Total spin of the system as a function of filling. Parameters: $L_x \times L_y= 20 \times2$, $t=1$, $D=2000$.}
\label{fig:DMRG_spin}
\end{figure}

To test whether the system is in the canted or ferromagetic state we computed the total spin $S_{tot}^2$ of the system as a function of filling in Fig. \ref{fig:DMRG_spin}. At half-filling $n=1$ and $U=16$, we observe $S^2=s_{max}(s_{max}+1)=420$ consistent with one spin per unit cell $s_{max}=(1/2) L_x L_y=20$. This result is also consistent with the original Lieb prediction of ferrimagnetism \cite{Lieb1989}. In the case of infinite interactions, we observe almost the full polarization when the system is doped with two holes away from half-filling. This polarization is associated with Nagaoka ferromagnetism.

Away from half-filling, the total spin rapidly goes to zero for both cases, signifying that the system is in the singlet state. The discrepancy between the DMRG and mean-field calculations could be explained by the fact that the mean-field usually overestimates the development of magnetically ordered phases. Furthermore, DMRG calculations are not truly 2 dimensional and $L_y>2$ analysis is required, which significantly complicates the calculations.

\section{Discussion}

This paper has introduced a $\mathbb{Z}_2$ fractionalized metallic state for the Lieb lattice Hubbard model at large Hubbard repulsion near filling fraction $\nu = 1/4$ which preserves all the symmetries of the lattice model. This is proposed as an attractive candidate to explain observations on ultracold atoms \cite{Lebrat24}, including the unexpected enhanced compressibility on the $p_{x,y}$ sites near $\nu = 1/4$. This fractionalized metal is obtained by doping a conventional ferrimagnetic insulator at $\nu = 1/2$, and so realizes the long-sought phenomenon of doping-induced fractionalization. 

The structure of the fractionalized insulator is most easily understood from the parton construction in Section~\ref{sec:sb}, which fractionalizes the spin-1/2 fermion $c_\sigma$ into a bosonic spinon $b_\sigma$ and a fermionic spinless `chargon' $f$. 
The nearest-neighbor exchange prefers a condensate of $Q_{ij} \sim \epsilon_{\sigma\sigma'}
b_{i \sigma} b_{j \sigma'}$, where $i,j$ are nearest-neighbors. On the other hand, the hopping term prefers a condensate of $P_{ij} \sim b_{i \sigma}^{\dagger} b_{j \sigma}$. The presence of both $Q_{ij}$ and $P_{ij}$ condensate leaves only the $\mathbb{Z}_2$ gauge symmetry $b_{i \sigma} \rightarrow \pm b_{i \sigma}$ unhiggsed, leading to the $\mathbb{Z}_2$ fractionalized state with the same anyon sectors as the toric code. The $b_{\sigma}$ excitations then represent the gapped spinon sector (the $e$ particles) of the fractionalized metal. When the $b_{\sigma}$ condense, all fractionalized excitations are confined in the state with canted magnetic order in Fig.~\ref{fig:canted}. In the fractionalized state, the chargons have effective filling fraction $\nu_{f} = 1-2 \nu$, and so $\nu_{f} = 1/2$ when $\nu = 1/4$. The chargon band structure is shown in Fig.~\ref{fig: holon bands}, and there is a flat $p_{x,y}$ flat band near the Fermi level at $\nu=1/4$, as hinted in Ref.~\cite{Lebrat24}.

Establishing the existence of $\mathbb{Z}_2$ fractionalized state in future experiments requires lower temperatures and the observation of a spin gap, along with the absence of a ferromagnetic moment present in the canted magnetic state. 
Recent experimental work \cite{Xu2025} has demonstrated progress towards achieving temperatures comparable to the expected spin gap.
The spiral states can be detected by measuring the susceptibility to Ising-nematic ordering.

The nearly flat $p_{x,y}$ band near the Fermi level in the $\mathbb{Z}_2$ fractionalized metal indicates that this phase is likely not stable down to zero temperature. Eventual instabilities to charge density wave order and superconductivity appear likely and are interesting topics for further study.

\section*{Acknowledgements}

P.M.B. thanks Z.~Han for valuable discussions.
This research was supported by the U.S. National Science Foundation grant No. DMR-2245246 and by the Simons Collaboration on Ultra-Quantum Matter which is a grant from the Simons Foundation (651440, S.S.). P.M.B. acknowledges support by the German National Academy of Sciences Leopoldina through Grant No.~LPDS 2023-06. A.K. acknowledges support from the NSF Graduate Research Fellowship Program. M.L. acknowledges support from the the Swiss National Science Foundation and the Max Planck/Harvard Research Center for Quantum Optics. 

\appendix

\bibliography{main}
\end{document}